\documentclass[a4paper,11pt]{article}
\pdfoutput=1
\usepackage{jheppub,multirow,enumitem}
\usepackage[table]{xcolor}
\usepackage{hhline} 
\usepackage{float} 
\usepackage[T1]{fontenc} 

\graphicspath{{./1_Figures/}}

\DeclareMathOperator\Tr{Tr}
\renewcommand{\bar}[1]{\overline{#1}}
\newcommand{\Or}{\mathcal{O}}
\newcommand{\bep}{\begin{pmatrix}} 
\newcommand{\eep}{\end{pmatrix}}
\newcommand{\U}{\text{U}}
\newcommand{\1}{1}
\newcommand{\0}{0}
\newcommand{\ZZ}{\mathbb{Z}}
\newcommand{\rmd}{d}
\newcommand{\rme}{e}
\newcommand\mequiv=
\newcommand\E{\lambda}
\newcommand{\NN}{{\cal N}}
\newcommand{\qq}{{\hat{q}}}
\newcommand{\VV}{\bar{V}}
\newcommand{\DD}{\bar{D}}

\def\ba#1\ea{\begin{align}#1\end{align}}
\def\mkakko#1{\left( #1 \right)}
\def\kkakko#1{\left[ #1 \right]}
\def\kakko#1{\langle #1 \rangle}
\def\wt#1{\widetilde{#1}}
\def\alert#1{\textcolor{red}{#1}}

\definecolor{cGOE}{rgb}{1,0.5,0.9}
\definecolor{cGSE}{rgb}{0.3,0.8,1}
\definecolor{cgray}{gray}{0.92}

\title{\boldmath Complete random matrix classification of 
SYK models with $\NN=0$, $1$ and $2$ supersymmetry}

\author[a]{Takuya Kanazawa}
\author[b]{and Tilo Wettig}
\affiliation[a]{iTHES Research Group and Quantum Hadron Physics Laboratory, 
RIKEN, 6-7-1 Minatojima-minamimachi, Chuo-ku, Kobe, Hyogo 650-0047, Japan}
\affiliation[b]{Department of Physics, University of Regensburg, 93040 Regensburg, Germany}
\emailAdd{takuya.kanazawa@riken.jp}
\emailAdd{tilo.wettig@ur.de}
\preprint{RIKEN-QHP-311}
\abstract{We present a complete symmetry classification of the Sachdev-Ye-Kitaev (SYK) model 
with $\NN=0$, 1 and 2 supersymmetry (SUSY) on the basis of the Altland-Zirnbauer 
scheme in random matrix theory (RMT). For $\NN=0$ and $1$ we consider 
generic $q$-body interactions in the Hamiltonian and find 
RMT classes that were not present in earlier classifications
of the same model with $q=4$. We numerically establish quantitative 
agreement between the distributions of the smallest energy levels in 
the $\NN=1$ SYK model and RMT. Furthermore, 
we delineate the distinctive structure of the $\NN=2$ SYK model and 
provide its complete symmetry classification based on RMT
for all eigenspaces of the fermion number operator. We corroborate our classification by 
detailed numerical comparisons with RMT and thus
establish the presence of quantum chaotic dynamics in the $\NN=2$ SYK model. 
We also introduce a new SYK-like model without SUSY that exhibits hybrid 
properties of the $\NN=1$ and $\NN=2$ SYK models 
and uncover its rich structure both analytically and numerically.
}

\begin{document} 
\maketitle
\flushbottom

\section{Introduction}

Understanding the mechanism of thermalization and information spreading 
(scrambling) in nonequilibrium quantum many-body systems 
is one of the fundamental challenges in theoretical physics. In a classically chaotic system  
the information on the initial conditions is quickly lost, which can be 
measured by the Lyapunov exponent that characterizes the sensitivity of 
the orbit to perturbations of initial conditions. In quantum systems, 
one clear fingerprint of chaos is the fact that statistical properties of the energy levels
are given by random matrix theory (RMT) \cite{Bohigas:1983er,Guhr:1997ve,
Muller2009NJP}. Quantum chaos in this sense 
has been the subject of research over decades, and its possible role in 
the relaxation (or thermalization) of a quantum system to equilibrium 
is still actively debated 
\cite{Deutsch1991,Srednicki1994,Biro:1994bi,Stockmann2007,
HaakeBook,Gomez2011review,GogolinEisert2016,DAlessio2016review}. 
Further progress was made on the  
treatment of black holes and holography in terms of quantum information theory 
\cite{Sekino:2008he,Sachdev:2010um,Susskind:2011ap,Lashkari:2011yi,Shenker:2013pqa,Shenker:2013yza,
Harlow:2014yka,Roberts:2014isa,Roberts:2014ifa,Shenker:2014cwa}. 
Building on these works, Kitaev suggested to employ 
the so-called out-of-time-ordered correlator (OTOC) \cite{Larkin1969} 
to probe information scrambling in black holes and in more 
general quantum systems \cite{KitaevTalks2014}. 
Along this line of thought one can define a quantum 
analog of the classical Lyapunov exponent, which is argued to have 
an intrinsic upper bound under certain assumptions 
\cite{Maldacena:2015waa}. Based on earlier work of Sachdev and Ye 
\cite{Sachdev:1992fk}, Kitaev put forward a $(0+1)$-dimensional fermionic model 
with all-to-all random interactions that can be solved in the large-$N$ 
limit, with $N$ the number of fermions involved \cite{KitaevTalks2015}. 
While it is hard to avoid the spin-glass phase at low temperatures
in the original Sachdev-Ye model \cite{Georges2000,Georges2001a}, it is 
ingeniously avoided in Kitaev's model, where fermions are put on a single site. 
Despite its apparent simplicity, this new Sachdev-Ye-Kitaev (SYK) model has 
a number of intriguing properties, including the spontaneous breaking 
of reparametrization invariance, emergent conformality at low energy, 
and maximal quantum chaos at strong coupling that points to 
an underlying duality to a black hole \cite{KitaevTalks2015,Sachdev:2015efa,
Polchinski:2016xgd,Maldacena:2016hyu,Maldacena:2016upp,Engelsoy:2016xyb}. 
Since the model was announced,  
a variety of generalizations appeared and computations 
of the OTOC in various other models were performed  
\cite{Hosur:2015ylk,Fu:2016yrv,Roberts:2016wdl,Jensen:2016pah,
Bagrets:2016cdf,Rozenbaum:2017zfo,Tsuji:2016jbo,Berkooz:2016cvq,Roberts:2016hpo,
Patel:2016wdy,Davison:2016ngz,Balasubramanian:2016ids,Liu:2016rdi,
Bagrets:2017pwq,Chowdhury:2017jzb,Hashimoto:2017oit}, 
including an SYK-like tensor model without random disorder \cite{Witten:2016iux}, 
modified SYK models with a tunable quantum 
phase transition to a nonchaotic phase \cite{Banerjee:2016ncu,Bi:2017yvx,Chen:2017dav}, 
and supersymmetric generalizations of the SYK model 
\cite{Fu:2016vas} (see also \cite{Anninos:2016szt,Gross:2016kjj,Sannomiya:2016mnj,
Yoon:2017gut,Peng:2017spg}). 
An analysis of tractable SYK-type models with SUSY will not only help to better understand 
theoretical underpinnings of the original AdS/CFT correspondence \cite{Maldacena:1997re} 
but also provide insights into condensed matter 
systems with emergent SUSY at low energy 
\cite{Ponte:2012ru,Grover:2013rc,Jian:2014pca,Rahmani:2015qpa,Jian:2016zll}. 

The level statistics of the SYK model was numerically examined in 
\cite{You:2016ldz,Garcia-Garcia:2016mno,Cotler:2016fpe,Garcia-Garcia:2017pzl} via 
exact diagonalization and agreement with RMT was found (although sizable 
discrepancies from RMT were seen in the long-range correlation \cite{Garcia-Garcia:2016mno}). 
An intimate connection between the SYK model and the so-called 
$k$-body embedded ensembles of random matrices \cite{Benet:2002br,Kota2014book} 
was also pointed out \cite{Garcia-Garcia:2016mno}. 
The algebraic symmetry classification of the SYK model based on RMT 
in \cite{You:2016ldz,Garcia-Garcia:2016mno,Cotler:2016fpe} was 
recently generalized to the $\NN=1$ supersymmetric SYK model \cite{Li:2017hdt}. 
A random matrix analysis of tensor models has also appeared 
\cite{Krishnan:2016bvg,Krishnan:2017ztz}. 

In this paper, we complete the random matrix analysis of the SYK model. Specifically:
\begin{enumerate}\itemsep0mm
	\item 
	We extend the symmetry classification of SYK models with $\NN=0$ and $1$ SUSY 
	that were focused on the 4-body interaction Hamiltonian 
	\cite{You:2016ldz,Garcia-Garcia:2016mno,Cotler:2016fpe} 
	to generic $q$-body interactions. The correctness of our classification is 
	then checked by detailed numerical simulations of the SYK model.  
	\item 
	We provide a detailed numerical examination of the hard-edge universality 
	of energy-level fluctuations near zero in SYK models.   
	\item 
	We delineate the complex structure of the Hilbert space of the $\NN=2$ SYK model 
	and provide a complete random matrix classification of energy-level statistics 
	in each eigenspace of the fermion number operator.  
\end{enumerate} 

This paper is organized as follows. In section~\ref{sc:rmt} we review 
the random matrix classification of generic Hamiltonians 
to make this paper self-contained. 
In section~\ref{sc:N0} we study the non-supersymmetric SYK model. We determine the relevant 
symmetry classes and report on detailed numerical verifications. 
In section~\ref{sc:N1SYK} we study the $\NN=1$ supersymmetric SYK model in a similar fashion. 
In section~\ref{sc:interlude} we introduce a new SYK-like model that shares 
some properties (e.g., numerous zero-energy ground states) with the $\NN=2$ 
SYK model but is theoretically much simpler. 
In section~\ref{sc:N2SYK} we investigate the $\NN=2$ supersymmetric SYK model. 
We explain why the symmetry classification of this model is far more complex 
than for its $\NN=1$ and $0$ cousins. We identify random matrix ensembles for 
each eigenspace of the fermion number operator and present a quantitative comparison 
between the level statistics of the model and RMT by exact diagonalization. 
Section~\ref{sc:conclusion} is devoted to a summary and conclusions. 

Throughout this paper, we will denote the number of Majorana fermions by $N_m$ 
and the number of complex fermions by $N_c$\,. The number of fermions in the Hamiltonian 
is denoted by $q$ and that in the supercharge is denoted by $\qq$. 
Needless to say, $q$ is even and $\qq$ is odd.

\section{\label{sc:rmt}\boldmath Symmetry classes in RMT}

To set the stage for our later discussion focused on the supersymmetric SYK model, we begin with 
a pedagogical summary of the symmetry classification scheme for a generic Hamiltonian, 
also known as the Altland-Zirnbauer theory \cite{Altland:1997zz,Zirnbauer:1996zz,Heinzner:2004xj}. 
For broad reviews of RMT we refer the reader to 
\cite{Beenakker:1997zz,Guhr:1997ve,Verbaarschot:2000dy,Mehta_book,
Fyodorov:2004ar,Edelman_Rao_2005,handbook:2010,AGZbook,Forrester_book,Beenakker:2014zza}. 

In the early days of RMT, there were just 3 symmetry classes called the Wigner-Dyson ensembles, 
which can be classified by the presence or absence of the time-reversal 
invariance and the spin-rotational invariance of the Hamiltonian 
\cite{Wigner1958,Dyson:1962one,Dyson:1962two,Dyson:1962three}. It is convenient to 
distinguish them by the so-called Dyson index $\beta$, which counts the number of degrees of freedom 
per matrix element in the corresponding random matrix ensembles: 
$\beta=1$, $2$, and $4$ corresponds, respectively, to 
the Gaussian Orthogonal Ensemble (GOE), 
the Gaussian Unitary Ensemble (GUE), and the Gaussian Symplectic Ensemble (GSE). 
By diagonalizing a random matrix drawn from each ensemble, one finds the joint probability 
density for all eigenvalues $\{\lambda_n\}$ to be of the form 
$P(\lambda) \propto \prod_{i<j}|\lambda_i-\lambda_j|^{\beta} 
\prod_{n}\rme^{-V(\lambda_n)}$, where $V(x)\propto x^2$ 
is a Gaussian potential. The spectral density $R(\lambda)$, also called the one-point 
function, measures the number of levels in a given interval $[\lambda,\lambda+\rmd\lambda]$. 
In RMT, one can show under mild assumptions that for large matrix dimension this function 
approaches a semicircle $R(\lambda)\propto \sqrt{\Lambda^2-\lambda^2}$ (Wigner's semicircle law), 
but in real physical systems $R(\lambda)$ is typically sensitive to the microscopic details of the Hamiltonian,
and one cannot exactly match $R(\lambda)$ in RMT with the physical spectral density. By contrast, if one looks into  
level correlations after ``unfolding'', which locally normalizes the level density to $1$, one encounters 
universal agreement of physical short-range spectral correlations with RMT.\footnote{A cautionary remark is in order. When unitary symmetries are present, the Hamiltonian can be transformed to a block-diagonal form, where  
each block is statistically independent. The spectral correlations must then be measured in each 
independent block. If one sloppily mixes up all eigenvalues before measuring the spectral correlations, the 
outcome is just Poisson statistics (see 
section~III.B.5 of \cite{Guhr:1997ve} for a detailed discussion).}
Heuristically, larger $\beta$ implies stronger level repulsion and a more rigid spectrum. 
A quantum harmonic oscillator exhibits a spectrum 
with strictly equal spacings, while a completely random point process allows two levels to come arbitrarily 
close to each other with nonzero probability. RMT predicts a nontrivial behavior that falls in 
between these two extremes. It is well known that a quantum system whose classical limit is chaotic
tends to exhibit energy-level statistics well described by RMT 
\cite{Bohigas:1983er,Stockmann2007,HaakeBook}. Also, Wigner-Dyson 
statistics emerges in mesoscopic systems with disorder, where the theoretical understanding was 
achieved by Efetov \cite{Efetov:1997fw}.

An important property of the $\beta=4$ class is the Kramers degeneracy of levels. 
In general, when there is an antiunitary operator $P$ acting on the Hilbert space 
that commutes with the Hamiltonian, $P^{-1}HP=H$, it follows that  
for each eigenstate $\psi$ there is another state $P\psi$ that has the same energy as $\psi$. 
If $P^2=\1$ (GOE), $P\psi$ is not necessarily linearly independent of $\psi$, hence 
levels are not degenerate in general, whereas if $P^2=-\1$ (GSE) their linear independence can 
be readily shown, so that all levels must be twofold degenerate. We note that the existence of 
such an operator is a sufficient, but not necessary, condition for the degeneracy of eigenvalues. 

Long after the early work by Wigner and Dyson, 7 new symmetry classes were identified 
in physics. Hence there are now 10 classes in total. (Some authors count them as 12 by 
distinguishing subclasses more carefully, as we will describe later.) The salient feature pertinent 
to those post-Dyson classes is a spectral mirror symmetry: the energy levels are symmetric 
about the origin (also called ``hard edge''). 
This means that, while they show the standard 
GUE/GOE/GSE level correlations in the bulk of the spectrum (i.e., sufficiently far 
away from the edges of the energy band), their level density exhibits a universal shape  
near the origin, different for each symmetry class. (Such a property is absent in the 
Wigner-Dyson classes since the spectrum is translationally invariant after unfolding 
and there is no special point in the spectrum.) The physical significance of such near-zero 
eigenvalues depends on the specific context in which RMT is used. In Quantum Chromodynamics 
(QCD), small eigenvalues of the Dirac operator in Euclidean spacetime are intimately 
connected to the spontaneous breaking of chiral symmetry and the origin of mass 
\cite{Banks:1979yr,Leutwyler:1992yt,Verbaarschot:2000dy}. 
In mesoscopic systems that are in proximity to superconductors, small energy levels describe 
low-energy quasiparticles and hence affect transport properties of the system 
at low temperatures. In supersymmetric theories the minimal energy is nonnegative, and 
it takes a positive value when SUSY is spontaneously broken 
\cite{Witten:1982df,Cooper:1994eh,Weinberg:2000cr}. 

The three \emph{chiral} ensembles \cite{Shuryak:1992pi,Verbaarschot:1993pm,
Verbaarschot:1994qf,Verbaarschot:1997bf,Verbaarschot:2000dy} 
relevant to systems with Dirac fermions 
such as QCD and graphene are denoted by chGUE/chGOE/chGSE 
(also known as the Wishart-Laguerre ensembles) and have the block structure 
\scalebox{0.8}{$\displaystyle \bep \0 & \fbox{$*$} \\ \fbox{$*$} & \0\eep$}, 
which anticommutes with the 
chirality operator \smash[t]{$\gamma_5=\scalebox{0.8}{$\displaystyle 
\bep \1 & \0 \\ \0 & -\1 \eep$}$}. This accounts for the spectral mirror symmetry in these 3 classes. We remark that 
chiral symmetry (i.e., a unitary operation that anticommutes with the Hamiltonian) 
is often called a sublattice symmetry in the condensed matter literature. A unique 
characteristic of the chiral classes in contrast to the other 4 mirror-symmetric classes
is that there can be an arbitrary number of exact zero modes. This is easily seen 
by making the matrix block $\fbox{$*$}$ rectangular, say, of size $m\times n$. 
When $|m-n|$ is large, the nonzero levels are pushed away from the origin due to 
level repulsion. In the limit $m,n\to\infty$ with $m/n \not\to 1$ the macroscopic spectral 
density fails to approach Wigner's semicircle and rather converges to what is called 
the Mar\v{c}enko-Pastur distribution \cite{MP1967}.  
In the thermodynamic limit of QCD with nonzero fermion mass, the number of 
zero modes $|m-n|\propto V_4^{1/2}$ \cite{Leutwyler:1992yt} 
while $m,n\propto V^{}_4$\,, where $V^{}_4$ is the Euclidean spacetime volume, 
and hence the physical limit is $m/n\to 1$. 

The other 4 post-Dyson classes are referred to as the Bogoliubov-de~Gennes (BdG) ensembles. 
They were identified by Altland and Zirnbauer \cite{Altland:1996zz,Altland:1997zz}. It is the particle-hole symmetry 
that accounts for the mirror symmetry of the spectra in these classes. This completes 
the ten-fold classification of RMT as summarized in table~\ref{tb:RMTlist}. There is a 
one-to-one correspondence between each ensemble and symmetric spaces in 
Cartan's classification, so the RMT ensembles are often called by abstract names such as 
A, AI, and AII due to Cartan \cite{Zirnbauer:1996zz}. 
In recent years this classification scheme was found to be useful in 
the classification of topological quantum materials \cite{Schnyder:2008tya,
Kitaev:2009mg,Hasan:2010xy,Ryu:2010zza,Beenakker:2014zza,Chiu2016review}. 

\begin{table}[t]
	\caption{\label{tb:RMTlist}Classification of RMT symmetry classes. 
	In the first three rows we list the Wigner-Dyson 
	classes. $\beta$ is the Dyson index defined in the main text. In the remaining rows we list 
	the chiral and BdG classes. The joint probability density for energy levels in these ensembles 
	assumes the form $P(\lambda) \propto 
	\prod_{i<j}|\lambda_i^2-\lambda_j^2|^{\beta}\prod_{n}|\lambda_n|^{\alpha}$, 
	and the indices $\beta$ and $\alpha$ are presented in the third and fourth column, respectively. 
	$\alpha$ is related to the number of exact zero modes. The index $\nu$ 
	defined in the last column is related to the topological charge of the gauge field in non-Abelian 
	gauge theories. Here we define $\nu$ to be nonnegative. The symbol ``---'' implies that there is no symmetry 
	in that class. The classes B and DIII-odd are sometimes omitted in other references, but we include them here
	for completeness. $T_+$ ($T_-$) denotes an antiunitary operator that commutes (anticommutes) with the Hamiltonian, 
	and $\Lambda$ is the chirality operator.  
	If both $T_+$ and $T_-$ are present, there is chiral symmetry, 
	but the converse is not true in general. Our notation in this table is such that 
	$\bar{A}$ is the complex conjugate of $A$ and $A^\dagger$ is 
	the conjugate transpose of $A$, i.e., $A^\dagger=\bar{A}^T$.}
	\vspace{10pt}
	\centering
	\scalebox{0.85}{\noindent 
	\renewcommand{\arraystretch}{1.2}
	\begin{tabular}{|c|c|c|c|c|c|c|c|c|c|c|}
          \hline
		RMT & \multicolumn{2}{c|}
		{$\begin{array}{c}\text{Cartan}\vspace{-7pt}\\
		\text{name}\end{array}$} 
		& $\beta$ & $\alpha$ & 
		$T_+^2$ & $T_-^2$ & 
		$\Lambda^2$ & 
		\multicolumn{2}{c|}{$\begin{matrix}
		\text{Block}\vspace{-7pt}\\
		\text{structure}\end{matrix}$} 
		& $\begin{matrix}
		\text{\#Zero}\vspace{-7pt}\\\text{~modes}\end{matrix}$  
		\\\hhline{|=|==|=|=|=|=|=|==|=|}
		GUE & \multicolumn{2}{c|}{A} 
		&2& --- &---&---&
		---& \multicolumn{2}{c|}
		{$H=H^\dagger$ complex} & 0
		\\\hline 
		GOE & \multicolumn{2}{c|}{AI} 
		&1& --- & $+1$ & --- &
		---& \multicolumn{2}{c|}{$H=H^T$ real} & 0
		\\\hline 
		GSE & \multicolumn{2}{c|}{AII} 
		&4& --- & $-1$ & --- &
		---& \multicolumn{2}{c|}
		{$H=H^\dagger$ quaternion} & 0
		\\\hhline{|=|==|=|=|=|=|=|==|=|}
		chGUE & \multicolumn{2}{c|}{AIII} 
		&2& $2\nu+1$ & --- & --- &1& 
		\multicolumn{2}{c|}
		{$\bep \0 & W \\ W^\dagger & \0 \eep$, 
		$\begin{array}{c}
			W:\,\text{complex}\vspace{-5pt}\\ n\times m
		\end{array}$} & 
		\multirow{2}{*}[-10pt]{$\begin{array}{c}|n-m| 
		\vspace{-4pt}\\ (\equiv\nu) \end{array}$}
		\\\cline{1-10}
		chGOE & \multicolumn{2}{c|}{BDI} 
		&1& $\nu$ &$+1$&$+1$&1& 
		\multicolumn{2}{c|}
		{$\bep \0 & W \\ W^T & \0 \eep$, 
		$\begin{array}{c}
			W:\,\text{real}\vspace{-5pt}\\ n\times m
		\end{array}$} & 
		\\\cline{1-11}
		chGSE & \multicolumn{2}{c|}{CII} 
		&4& $4\nu+3$ & $-1$ & $-1$ &1& 
		\multicolumn{2}{c|}
		{$\bep \0 & W \\ W^\dagger & \0 \eep$, 
		$\begin{array}{c}
			W:\,\text{quaternion}\vspace{-5pt}\\ n\times m
		\end{array}$} & $2\nu$
		\\\hhline{|=|==|=|=|=|=|=|==|=|}
		\multirow{6}{*}[0pt]{\raisebox{-35pt}{BdG}} & 
		\multicolumn{2}{c|}{C} 
		&2&2&---& $-1$ &---& 
		\multicolumn{2}{c|}
		{$\bep A & B \\ \bar{B} & -\bar{A} \eep$, 
		$\begin{array}{l}
			\text{$A$ : Hermitian,}\\
			\text{$B$ : complex symmetric}
		\end{array}$}
		& 0
		\\\cline{2-11} 
		& \multicolumn{2}{c|}{CI} 
		&1&1&$+1$&$-1$&1& 
		\multicolumn{2}{c|}
		{$\bep \0 & Z \\ \bar{Z} & \0\eep$, 
		$Z$ : complex symmetric} 
		& 0
		\\\cline{2-11} 
		& \multirow{2}{*}[0pt]{BD} & D 
		&\multirow{2}{*}[0pt]{2}&0& 
		\multirow{2}{*}[0pt]{---}
		& \multirow{2}{*}[0pt]{$+1$} 
		& \multirow{2}{*}[0pt]{---} & 
		\multirow{2}{*}[2pt]{$\begin{array}{c}
			H \vspace{-8pt}\\ \text{\small pure imaginary} 
			\vspace{-6pt}\\ 
			\text{\small and skew-symmetric}
		\end{array}$} & dim$[H]=$ even & 0
		\\\cline{3-3}\cline{5-5}\cline{10-11} 
		& & B &&2&&&&&
		dim$[H]=$ odd & 1
		\\\cline{2-11} 
		& \multirow{2}{*}[-7pt]{DIII} & 
		$\begin{array}{cc}\text{DIII}
		\vspace{-7pt}\\\text{even}\end{array}$  
		& \multirow{2}{*}[-7pt]{4} &1&
		\multirow{2}{*}[-7pt]{$-1$}&
		\multirow{2}{*}[-7pt]{$+1$} & 
		\multirow{2}{*}[-7pt]{1} 
		&\multirow{2}{*}[4pt]{
		$\begin{array}{c}
			\bep \0 & Z \\ -\bar{Z} & \0\eep ,
			\vspace{-5pt}\\
			\text{$Z$ : \small complex and}
			\vspace{-8pt}\\
			\text{\small skew-symmetric}
		\end{array}$}& dim$[Z]=$ even & 0
		\\\cline{3-3}\cline{5-5}\cline{10-11} 
		& & $\begin{array}{cc}\text{DIII}
		\vspace{-7pt}\\\text{odd}\end{array}$
		&&5&&&&& dim$[Z]=$ odd & 2
		\\\hline
	\end{tabular}}
\end{table}

We refer the reader to \cite{Altland:1997zz,Zirnbauer:1996zz,Caselle:2003qa,
Heinzner:2004xj,Zirnbauer2004_review} 
for the detailed mathematics of the Altland-Zirnbauer theory and
only recall the essential ingredients here. Let $T_+$ ($T_-$) denote an antiunitary operator that 
commutes (anticommutes) with the Hamiltonian.%
\footnote{Here we conform to the notation of \cite{Li:2017hdt}. Rather than calling $T_\pm$  
time-reversal symmetry or spin-rotational symmetry, we prefer to denote them by abstract symbols, 
since the proper physical interpretation of each operator depends on the specific system.} 
(Note that any antiunitary operator can be expressed as the product 
of a unitary operator and the complex conjugation operator $K$.) The chirality operator (a unitary operator 
that anticommutes with the Hamiltonian and squares to $\1$) is denoted by $\Lambda$ from here on.
The first step is to check whether $T_+$, $T_-$, and $\Lambda$ 
exist for a given Hamiltonian. 
If both $T_+$ and $T_-$ exist, one always has chiral symmetry, $\Lambda=T_+T_-$. 
The second step is to check if the antiunitary symmetry
squares to $+\1$ or $-\1$. This allows one to figure out which class the Hamiltonian belongs to. 
However, there is an additional subtlety in the symmetry classes BD and DIII. There one has to distinguish two cases 
according to the parity of the dimension of the Hilbert space (see table~\ref{tb:RMTlist}), 
which results in the presence/absence 
of exact zero modes. The classes B and DIII-odd have physical applications to 
superconductors with $p$-wave pairing \cite{Bocquet2000,Ivanov1999,Ivanov2001,Ivanov2002super}. 
The functional forms of the universal level density near zero for all the 7 post-Dyson classes 
are explicitly tabulated in, e.g., \cite{Ivanov2002super,Gnutzmann2006review}. 
Note that, because class B and class C share the same set of 
indices $\alpha$ and $\beta$, 
their level density near zero coincides, except for a delta function 
at the origin in class B. 

\section{\label{sc:N0}\boldmath $\NN=0$ SYK model}

In this and the next section, we complete the random matrix 
classification of the SYK model with $\NN=0$ and $1$ SUSY with $q$-body 
interactions, generalizing earlier work focused mostly on $q=4$ 
\cite{You:2016ldz,Garcia-Garcia:2016mno,Cotler:2016fpe,Li:2017hdt}.  
Many of the concepts and techniques employed here will be taken up 
again for the analysis of the $\NN=2$ SYK model in section~\ref{sc:N2SYK}.

\subsection{Definitions of relevant operators\label{sc:PHT}}

To begin with, recall that when we speak of a non-SUSY SYK model, there are actually two models, 
one involving Majorana fermions \cite{KitaevTalks2015,Polchinski:2016xgd,Maldacena:2016hyu} 
and another involving complex fermions \cite{Sachdev:2015efa,
You:2016ldz,Fu:2016yrv,Davison:2016ngz,Bulycheva2017}. In either case 
it is useful to start with the creation and annihilation operators of 
\emph{complex} fermions, denoted by $\bar{c}_a$ and $c_a$, respectively, obeying 
\ba
	\{c_a, c_b\} = \{\bar{c}_a, \bar{c}_b\}=0\,, \quad 
	\{c_a, \bar{c}_b\}=\delta_{ab} 
	\quad \text{with } a=1,\dots,N_c \,. 
\ea
These operators can be represented as \emph{real} matrices by 
adopting the Jordan-Wigner construction \cite{Fu:2016yrv,You:2016ldz} 
$c_a=(\prod_{1\leq b<a}\sigma_b^z)(\sigma_a^x+i\sigma_a^y)/2$ 
and $\bar{c}_a = (c_a)^\dagger$.%
\footnote{The structure of energy levels including degeneracy is 
of course independent of the basis choice, but making $\bar{c}$ 
and $c$ real makes symmetry classification based on antiunitary 
operations more transparent.} 
We also define the fermion number operator
\ba
	F\equiv \sum_{a=1}^{N_c}\bar{c}_a c_a\,.
	\label{eq:Fdef}
\ea
The total Hilbert space $V$ of dimension $2^{N_c}$ 
splits into two sectors with even/odd eigenvalue of $F$, i.e., $(-1)^F=\pm 1$. 

One can construct $N_m=2N_c$ Majorana fermions $\chi_i$ from complex fermions as
\ba
	\chi_{2k-1} = \frac{c_{k} + \bar{c}_{k}}{\sqrt{2}}\,, 
	\quad 
	\chi_{2k} = \frac{c_{k} - \bar{c}_{k}}{\sqrt{2}\,i}\,, 
	\quad k=1,\dots, N_c\,, \quad 
	\{\chi_i,\chi_j\}=\delta_{ij}\,.
\ea
The antiunitary operator of special importance in the SYK model 
is the particle-hole operator \cite{Fidkowski2010,You:2016ldz,
Garcia-Garcia:2016mno,Cotler:2016fpe}
\ba
	P = K \prod_{a=1}^{N_c}(c_a+\bar{c}_a) 
	\equiv K (c_1+\bar{c}_1)\cdots (c^{}_{N_c}+\bar{c}^{}_{N_c})\,,
	\label{eq:defP}
\ea
where $K$ is complex conjugation. 
One can show \cite{You:2016ldz,Garcia-Garcia:2016mno,Cotler:2016fpe}
\begin{gather}
	P c_a P = \eta \bar{c}_a\,, \quad P \bar{c}_a P = \eta c_a\,, \quad 
	P \chi_i P = \eta \chi_i\,,
	\\
	P^2 = (-1)^{\lfloor N_c/2 \rfloor}\,, 
	\qquad \eta=(-1)^{\lfloor 
	\frac{N_c-1}{2}\rfloor}\,.
      \label{eq:eta}
\end{gather}
Here $\lfloor x \rfloor$ denotes the greatest integer that does not exceed $x$. 
We stress that all of the above formulas hold irrespective of the form of the Hamiltonian. 

\subsection{Classification} 

Let us begin with the non-supersymmetric SYK model with $N_m$ 
Majorana fermions for $N_m$ even.%
\footnote{The Hilbert space for odd $N_m$ can be 
constructed by adding another Majorana fermion that does not 
interact with the rest. For the symmetry classification of the SYK model with 
odd $N_m$, see \cite{You:2016ldz}.} 
For a positive even integer $2\leq q\leq N_m$, the Hamiltonian  
\cite{KitaevTalks2015,Polchinski:2016xgd,Maldacena:2016hyu} is given by
\ba
	H = i^{q/2} \sum_{1\leq i_1<\cdots <i_q \leq N_m}
	J_{i_1\cdots i_q}\chi_{i_1} \chi_{i_2} \cdots \chi_{i_q}\,,
	\label{eq:SYKm0}
\ea 
where $J_{i_1\cdots i_q}$ are independent real Gaussian random variables 
with the dimension of energy, $\langle J_{i_1\cdots i_q} \rangle=0$ 
and $\langle J_{i_1\cdots i_q}^2 \rangle = \frac{(q-1)!}{N_m^{q-1}}J^2$. 
The prefactor $i^{q/2}$ is necessary to make $H$ Hermitian. 
This model is conjectured to be dual to a black hole in the large-$N$ limit 
\cite{KitaevTalks2015,Polchinski:2016xgd,Maldacena:2016hyu} and for 
$\beta J\gg 1$ saturates the bound on quantum chaos proposed in \cite{Maldacena:2015waa}. 
While the $q=4$ version has attracted most of the attention in the literature, 
it is useful to consider general $q$ because the theory simplifies 
in the large-$q$ limit \cite{KitaevTalks2015,Maldacena:2016hyu}. 

Now, due to the Majorana nature of the fermions, 
the fermion number is only conserved modulo 2. The 
Hilbert space naturally admits a decomposition into two sectors of equal dimensions,  
with a definitive parity of the fermion number. Since $H$ does not mix sectors 
with $(-1)^F=+1$ and $-1$, $H$ acquires a block-diagonal form \scalebox{0.8}
{$\displaystyle \bep A & \0 \\ \0 & B \eep$}, 
where $A$ and $B$ are Hermitian square matrices of equal dimensions. 
By examining the commutation relation of $H$ and $P$, one finds that $q\mequiv 0$ (mod~4) and 
$q\mequiv 2$ (mod~4) have to be treated separately because 
$HP=(-1)^{q/2}PH$. The spectral statistics for $q\mequiv 2$ (mod~4)  
did not receive attention in \cite{You:2016ldz,
Garcia-Garcia:2016mno,Cotler:2016fpe,Li:2017hdt},%
\footnote{An exception is the simplest case $q=2$, which was 
analytically solved at finite $N_m$ \cite{Gross:2016kjj} 
and in the limit $N_m\to\infty$ \cite{Anninos:2016szt,Maldacena:2016hyu} 
(see also \cite{Magan:2015yoa,Cotler:2016fpe}). 
Note that $H$ in this theory is just a random mass with no interactions, so 
one cannot extrapolate features of $q=2$ 
to the more nontrivial $q\geq 4$ cases.} and we shall work it out below. 
This is a new result.

\paragraph{\boldmath $q\mequiv 0$ (mod~4)}\ \\
        In this case $[H,P]=0$. Thus $P$ corresponds to $T_+$ in table~\ref{tb:RMTlist}.
	For $N_m\mequiv 0$ and 4 (mod~8), $P$ is a bosonic operator and maps each 
	parity sector onto itself. For $N_m\mequiv 0$ (mod~8), $P^2=+1$ so that $H=\text{GOE}\oplus\text{GOE}$. 
	For $N_m\mequiv 4$ (mod~8), $P^2=-1$ so that $H=\text{GSE}\oplus\text{GSE}$. In both cases the two 
	blocks of $H$ are independent in general. Finally, for $N_m\mequiv 2$ and $6$ (mod~8) 
	$P$ is a fermionic operator and exchanges the two sectors. Hence $H=
	\scalebox{0.8}{$\displaystyle \bep A & \0 \\ \0 & \bar{A} \eep$}$, where $A=A^\dagger$ belongs to GUE. 
	It follows that the eigenvalues are twofold degenerate for $N_m\mequiv 2$, $4$ and 
	$6$ (mod~8), and unpaired only 
	for $N_m\mequiv 0$ (mod~8). 
	This is summarized in table~\ref{tb:rmtSYK00}, which is  
	consistent with \cite{You:2016ldz,Garcia-Garcia:2016mno,
	Cotler:2016fpe,Li:2017hdt}.

\paragraph{\boldmath $q\mequiv 2$ (mod~4)}\ \\
        Now $\{H,P\}=0$. 
	Thus $P$ corresponds to $T_-$ in table~\ref{tb:RMTlist} and 
	the spectrum enjoys a mirror symmetry $\E\leftrightarrow -\E$.%
	\footnote{What is meant here is that 
	the mirror symmetry is present for \emph{every} single realization
	$\{J_{i_1,\cdots,i_q}\}$ of the disorder.}
	For $N_m\mequiv 0$ and 4 (mod~8), $P$ is a bosonic operator and maps each 
	parity sector onto itself. For $N_m\mequiv 0$ (mod~8), 
	$P^2=+1$ so that $H=\text{BdG(D)}\oplus\text{BdG(D)}$. 
	(It is not class B because the dimension $2^{N_m/2-1}$ of each sector is even.) 
	For $N_m\mequiv 4$ (mod~8), $P^2=-1$ so that $H=\text{BdG(C)}\oplus\text{BdG(C)}$. In both cases the two 
	blocks of $H$ are independent in general. For $N_m\mequiv 2$ and $6$ (mod~8), $H= 
	\scalebox{0.8}{$\displaystyle \bep A & \0 \\ \0 & -\bar{A} \eep$}$, where $A=A^\dagger$ belongs to GUE, 
	for the same reason as above. This is summarized in table~\ref{tb:rmtSYK01}.

\bigskip

\begin{table}[h]
	\caption{\label{tb:rmtSYK00}
	Symmetry classification of $H$ in the Majorana SYK model (no SUSY) 
	for $q\mequiv 0$ (mod~4). This table is consistent with 
	\cite{You:2016ldz,Garcia-Garcia:2016mno,Cotler:2016fpe,Li:2017hdt}.}
	\vspace{9pt}
	\renewcommand{\arraystretch}{1.05}
	\centering 
	\scalebox{0.92}{
	\begin{tabular}{|c||c|c|c|c|}
		\hline 
		\cellcolor{cgray}\!\!\!\!
		$\begin{array}{c}\text{\footnotesize $\NN=0$ SYK}
		\vspace{-4pt}\\ 
		{\footnotesize q\mequiv 0~\text{(mod~4)}}  \end{array}$\!\!\!\!
		& \footnotesize{Block structure} 
		& \footnotesize{\!\!\!degeneracy\!\!\!}  & $\beta$ & 
		$\begin{array}{c}\text{\footnotesize mirror}\vspace{-8pt}\\
		\text{\footnotesize \!\!\!\!symmetry\!\!\!\!}\end{array}$ 
		\\\hline 
		\!$N_m\mequiv 0$ (mod~8)\!\! & $\bep A & \0 
		\vspace{-4pt}\\ \0 & B\eep$, 
		$\begin{array}{c}\text{$A,B$: real}
		\vspace{-4pt}\\\text{symmetric}\end{array}$ 
		& 1 & 1 & \multirow{4}{*}[-22pt]{\text{No}}
                \\\cline{1-4}
		\!$N_m\mequiv 2$ (mod~8)\!\! & 
		$\bep A & \0 \vspace{-4pt}\\ \0 & \bar{A} \eep$, $A$: Hermitian
		& 2 & 2 & 
		\\\cline{1-4}
		\!$N_m\mequiv 4$ (mod~8)\!\! & 
		$\bep A & \0 
		\vspace{-4pt}\\ \0 & B\eep$, 
		$\begin{array}{c}\text{$A,B$:}
		\vspace{-5pt}\\\text{quaternion}
		\vspace{-5pt}\\\text{real}
		\end{array}$ 
		& 2 & 4 & 
		\vspace{-3pt}\\\cline{1-4}
		\!$N_m\mequiv 6$ (mod~8)\!\! & 
		$\bep A & \0 \vspace{-4pt}\\ 
		\0 & \bar{A} \eep$, $A$: Hermitian
		& 2 & 2 & 
		\\\hline
	\end{tabular}}
    \end{table}
\begin{table}[h]
	\caption{\label{tb:rmtSYK01}
	Symmetry classification of $H$ in the Majorana SYK model (no SUSY) 
	for $q\mequiv 2$ (mod~4). For the block structure of each class 
	we refer to table~\ref{tb:RMTlist}.}
	\vspace{9pt}
	\centering 
	\scalebox{0.92}{
	\begin{tabular}{|c||c|c|c|c|}
		\hline 
		\cellcolor{cgray}
		\!\!\!\! 
		$\begin{array}{c}\text{\footnotesize $\NN=0$ SYK} 
		\vspace{-4pt}\\ 
		{\footnotesize q\mequiv 2~\text{(mod~4)}} \end{array}$ \!\!\!\! 
		& \footnotesize{Block structure}  
		& \footnotesize{\!\!\!degeneracy\!\!\!}  & $\beta$ & 
		$\begin{array}{c}\text{\footnotesize mirror}\vspace{-8pt}\\
		\text{\footnotesize \!\!\!\!symmetry\!\!\!\!}\end{array}$ 
		\\\hline \noindent\!$N_m\mequiv 0$ (mod~8) \!\!\!\! & $\bep A & \0 
		\vspace{-4pt}\\ \0 & B\eep$, 
		$A,B\in$ BdG(D) 
		& \multirow{4}{*}[-15pt]{1} 
		& \multirow{4}{*}[-15pt]{2} 
		& \multirow{4}{*}[-15pt]{\text{Yes}}
		\\\cline{1-2} 
		\!\!\!$N_m\mequiv 2$ (mod~8)\!\!\!\! & 
		$\bep A & \0 \vspace{-4pt}\\ 
		\0 & -\bar{A} \eep$, $A$: Hermitian
		&  &  & 
                \\\cline{1-2} 
		\!\!\!$N_m\mequiv 4$ (mod~8)\!\!\!\! & 
		$\bep A & \0 
		\vspace{-4pt}\\ \0 & B\eep$, 
		$A,B \in$ BdG(C)
		&  &  &    
		\\\cline{1-2}
		\!\!\!$N_m\mequiv 6$ (mod~8)\!\!\!\! & 
		$\bep A & \0 \vspace{-4pt}\\ 
		\0 & -\bar{A} \eep$, $A$: Hermitian
		&  &  &  
		\\\hline
	\end{tabular}}
\end{table}

As a generalization one can also consider a Hamiltonian 
that includes both a $q\mequiv 0$ (mod~4) term and a $q\mequiv 2$ (mod~4) 
term. Then $H$ has no antiunitary symmetry and the result is just $\text{GUE}\oplus\text{GUE}$, 
i.e., $H=\scalebox{0.8}{$\displaystyle \bep A & \0 \\ \0 & B \eep$}$ with $A$ and $B$  
independent Hermitian matrices.

Even when the symmetry class of $H$ is known, it is 
highly nontrivial whether the level correlations of $H$ 
quantitatively coincide with those of RMT. In the SYK model 
\eqref{eq:SYKm0} there are only $\Or(N_m^q)$ 
independent random couplings, while a dense random matrix 
has $\Or(2^{N_m})$ independent random elements. 
The level statistics of $H$ for $q=4$ has been studied numerically via 
exact diagonalization \cite{You:2016ldz,Garcia-Garcia:2016mno,
Cotler:2016fpe,Garcia-Garcia:2017pzl} and agreement 
with the RMT classes in table~\ref{tb:rmtSYK00} 
was found for not too small $N_m$\,. This is consistent 
with the quantum chaotic behavior of the model 
\cite{KitaevTalks2015,Maldacena:2016hyu}. 

\subsection{Numerical simulations}

\begin{figure}[t]
	\centering
	\includegraphics{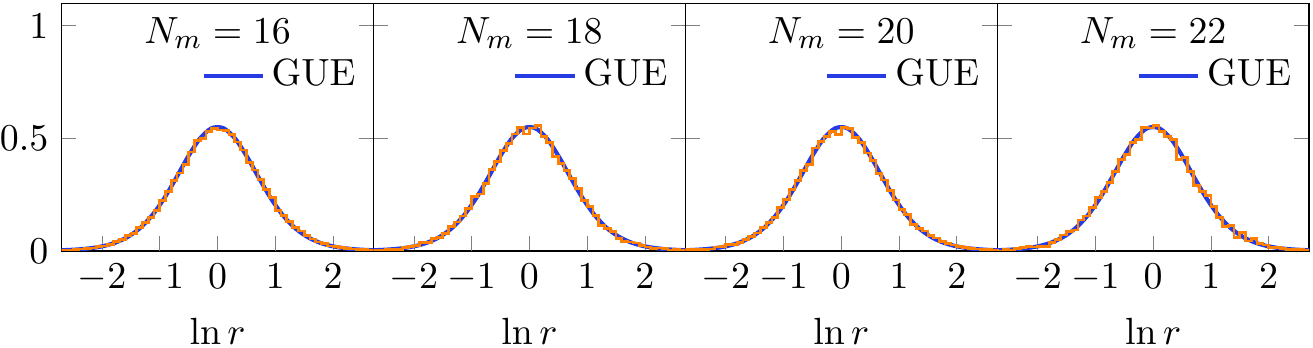}
	\vspace{-8pt}
	\caption{\label{fg:N0q6logr}Statistical distribution of the ratio $r$ of 
	two consecutive level spacings for the $\NN=0$ SYK model with $q=6$. 
	The number of realizations used for averaging was $10^3$ for $N_m=16$, 
	$10^2$ for $N_m=18$ and $20$, and 10 for $N_m=22$. 
	The blue lines are surmises for the RMT classes in table~\ref{tb:rmtSYK01}.}
\end{figure}

\paragraph{Level correlations in the bulk} \ \\
Here we report on the first numerical analysis of the bulk statistics of energy levels for the 
\mbox{$\NN=0$} SYK model with $q=6$ via exact diagonalization to test table~\ref{tb:rmtSYK01}. 
To identify the symmetry class we employ the probability distribution $P(r)$
of the ratio 
$r=(\lambda_{n+2}-\lambda_{n+1})/(\lambda_{n+1}-\lambda_{n})$ 
of two consecutive level spacings in a sorted spectrum, 
as it does not require an unfolding procedure 
\cite{Oganesyan2007,Atas2012PRL,You:2016ldz}. We used accurate 
Wigner-like surmises for the Wigner-Dyson classes derived in \cite{Atas2012PRL},
\begin{align}
  P_{\mbox{\tiny$W$}}(r) & = \frac{1}{Z_\beta}\frac{(r+r^2)^\beta}
                           {(1+r+r^2)^{1+3\beta/2}}
\end{align}
with $Z_1=8/27$, $Z_2=4\pi/81\sqrt3$, and $Z_4=4\pi/729\sqrt3$.
For Poisson statistics we have $P_{\mbox{\tiny$P$}}(r)=1/(1+r)^2$ \cite{Atas2012PRL}.
Our numerical results are displayed in figure~\ref{fg:N0q6logr}. 
Without any fitting parameter, they all agree excellently with the GUE $(\beta=2)$ 
as predicted by table~\ref{tb:rmtSYK01}. 
This indicates that quantum chaotic dynamics emerges in this model even for
such small values of $N_m$.

\begin{figure}[h]
	\centering
	\includegraphics{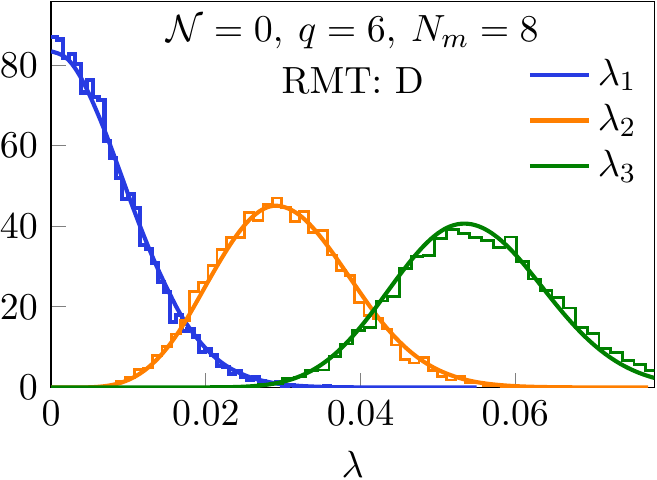}
	\qquad
	\includegraphics{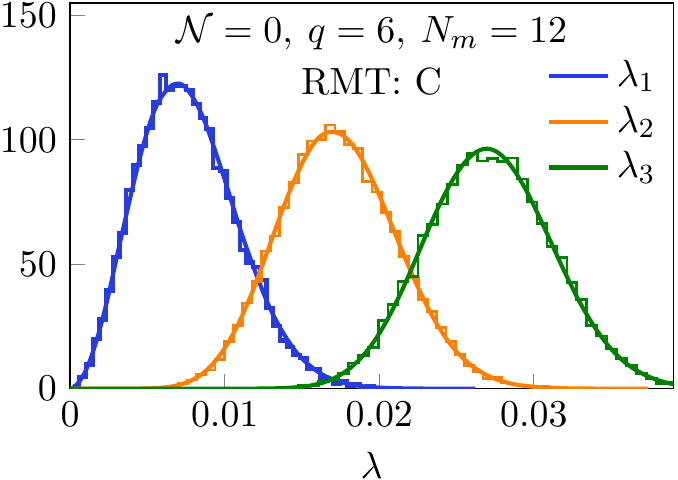}
	\vspace{-7pt}
	\caption{\label{fg:PminN0}Distributions of the eigenvalues 
	of $H$ with smallest absolute values in the $\NN=0$ SYK model 
	with $q=6$ and $J=1$, 
	compared with the predictions (solid lines) of the RMT 
	classes in table~\ref{tb:rmtSYK01}. 
	The number of independent random samples is $10^4$ for each plot. The small deviations from RMT for the third nonzero eigenvalue are interpreted to be effects of finite $N_m$.}
\end{figure}

\paragraph{Universality at the hard edge} \ \\
In class C and D the origin is a special point due to the spectral mirror symmetry, 
and the level statistics near zero shows universal fluctuations different from those in 
the bulk of the spectrum \cite{Altland:1997zz}. Their form is solely determined 
by the global symmetries of the Hamiltonian and is insensitive to microscopic details of 
interactions. In figure~\ref{fg:PminN0} we compare the distributions of the 
near-zero energy levels of the $\NN=0$ SYK model with $q=6$  
and those of RMT, finding nearly perfect agreement.%
\footnote{To obtain these plots we determined the RMT curves
  numerically for matrix size $10^3$ using the mapping to tridiagonal
  matrices invented in \cite{Dumitriu-Edelman}. We then rescaled the
  RMT curves as $p(x)\to c p(cx)$ and tuned the parameter $c$ to
  achieve the best fit to the data, where $c$ is common to the three
  curves in each plot.}
The nonzero (zero) 
intercept at $\lambda=0$ in class D (class C) directly reflects 
the fact that $\alpha=0$ for class D ($\alpha=2$ for class C), 
where $\alpha$ is the index listed in 
table~\ref{tb:RMTlist}.

\subsection[Overview of the $\NN=0$ SYK model with complex fermions]
{\label{sc:compsyk}\boldmath Overview of the $\NN=0$ SYK model with complex fermions}

We finally comment on the non-supersymmetric 
SYK model with complex fermions \cite{Sachdev:2015efa,
You:2016ldz,Fu:2016yrv,Davison:2016ngz,Bulycheva2017}. 
The Hamiltonian reads 
$H=\sum_{i,j,k,\ell=1}^{N_c}J_{ij;k\ell}
\bar{c}_i\bar{c}_jc_kc_\ell - \mu F$, where $\mu$ is 
the chemical potential for the fermion number operator $F$ in \eqref{eq:Fdef}
and the coupling is a complex Gaussian random variable obeying 
$J_{ij;k\ell}=-J_{ji;k\ell}=-J_{ij;\ell k}=J_{k\ell;ij}^{*}$. 
Since $H$ preserves the fermion number, $H$ as a matrix 
has a block-diagonal structure representing 
each eigenspace of $F=0,1,\dots,N_c$\,. There is no antiunitary 
symmetry for $H$ and consequently the levels collected in 
each block of $H$ would obey GUE. Intriguingly, one can amend $H$ by adding 
correction terms so that it commutes with $P$ \cite{You:2016ldz,Fu:2016yrv}. 
In this case, the half-filled sector $F=N_c/2$ (which only exists for 
$N_c$ even) is symmetric under $P$ and its level statistics becomes 
either GOE (if $P^2=+1$) or GSE (if $P^2=-1$). In all other 
sectors, the level statistics remains GUE, but there arises a degeneracy 
between the sector $F=k$ and the sector $F=N_c-k$ 
for $k\ne N_c/2$ since they are mapped to each other by $P$.

\section{\label{sc:N1SYK}\boldmath $\NN=1$ SYK model}

\subsection{Classification}

The supersymmetric generalization of the SYK model 
was introduced in \cite{Fu:2016vas} 
(see also \cite{Anninos:2016szt,Gross:2016kjj,Sannomiya:2016mnj,Yoon:2017gut,Peng:2017spg}).  
The model with $\NN=1$ SUSY has the Hamiltonian $H=Q^2$ with supercharge 
\ba
	\label{eq:QdefN1}
	Q = i^{(\qq-1)/2} \!\!\! 
	\sum_{1\leq i_1<\cdots<i_{\qq}\leq N_m} 
	C_{i_1 i_2 \cdots i_{\qq}} \chi_{i_1} \chi_{i_2} \cdots \chi_{i_\qq}\,, 
\ea
where $1\leq \qq\leq N_m$ is an odd integer. (Note that $Q^\dagger=Q$.) 
In this case $H$ involves terms with
up to $2\qq-2$ fermions. The couplings $C_{i_1 i_2 \cdots i_{\qq}}$ are 
independent real Gaussian variables with mean 
$\kakko{C_{i_1 i_2 \cdots i_{\qq}}} = 0$ and variance 
$\kakko{C^2_{i_1 i_2 \cdots i_{\qq}}} = 
\frac{(\qq-1)!}{N_m^{\qq-1}}J$ for some $J>0$\,. 
The ground-state energy of this model is evidently nonnegative. 
In \cite{Fu:2016vas} a strictly positive ground-state energy that decreases exponentially 
with $N$ was obtained numerically, indicating that SUSY is dynamically 
broken at finite $N$ and restored only in the large-$N$ limit. 

It is easy to verify the simple relation
\begin{equation}
	\rho_H(\E) = \frac{1}{\sqrt{\E}} \rho_{Q}(\sqrt{\E}\,) \quad \qquad (\E\geq 0)
	\label{eq:HQrelate}
\end{equation}
between the spectral densities of 
$H$ and $Q$, where $\rho_H(\E) \equiv \big\langle \! \Tr \delta(\E - H) \big\rangle$ and 
$\rho_Q(X) \equiv \big\langle \! \Tr \delta(X - Q) \big\rangle$\,.  
Equation~\eqref{eq:HQrelate} reveals that the level density of 
$H$ would blow up as $\E^{-1/2}$ near zero if $Q$ had a nonzero density of states 
at the origin for large $N_m$. This blow-up was indeed seen in the 
exact diagonalization analysis \cite{Li:2017hdt} as well as in analytical studies of 
the low-energy Schwarzian theory \cite{Fu:2016vas,Stanford:2017thb,Mertens:2017mtv}. 
Since $Q$ is more fundamental than $H$ we will focus on the 
level structure of $Q$ below, viewing it as a matrix acting on the many-body 
Fock space. 

The random matrix classification for $\qq=3$ has recently been
put forward in \cite{Li:2017hdt}. Here we will generalize this 
to all odd $\qq$, with emphasis on the difference of symmetry classes between 
$\qq\mequiv 1$ (mod~4) and $\qq\mequiv 3$ (mod~4). 
The main theoretical novelty in the $\NN=1$ SYK model 
is the fact that $Q$ anticommutes with the fermion parity operator $(-1)^F$. 
Thus $(-1)^F$ plays the role of $\gamma_5$ for the Dirac operator in QCD 
and naturally induces a block structure 
\scalebox{0.8}{$\displaystyle \bep \0 & \fbox{$*$} \\ \fbox{$*$}^\dagger & \0 \eep$} for $Q$. 
The spectrum of $Q$ is therefore symmetric under $\E\leftrightarrow -\E$.  
Since the block $\fbox{$*$}$ is a square matrix, there are no 
topological zero modes, i.e., all eigenvalues of $Q$ are nonzero 
unless fine-tuning of the matrix elements is performed. 
From the relation $H=Q^2$ 
we conclude that all eigenvalues of $H$ should be at least \emph{twofold 
degenerate}. 
 
Following \cite{Li:2017hdt} we introduce a new antiunitary operator 
$R\equiv P(-1)^F$.  We have
\ba
	PQP = (-1)^{(\qq-1)/2} \eta Q  \qquad 
	\text{and} \qquad 
	RQR = (-1)^{(\qq-1)/2+N_c+1} \eta Q\,,
	\label{eq:PRcoms}
\ea
where $N_c=N_m/2$ as before and $\eta$ is given in \eqref{eq:eta}. These relations, combined with 
table~\ref{tb:RMTlist}, lead to the classification of $Q$ shown in 
table~\ref{tb:RMTQ00} for $\qq\mequiv 1$ (mod~4) 
and table~\ref{tb:RMTQ01} for $\qq\mequiv 3$ (mod~4). By comparing 
the (anti-)commutators in each table, we see that the roles of $P$ and $R$ 
are exchanged for $\qq\mequiv 1$ and 3. Consequently 
the positions of BdG(CI) and BdG(DIII-even) are exchanged. 
In these tables we made it clear that we are considering 
chGOE and chGSE in the topologically trivial sector $\nu=0$.  

\begin{table}[tb]
	\caption{\label{tb:RMTQ00}
	Symmetry classification of $Q$ in the $\NN=1$ SYK model  
	for $\qq\mequiv 1$ (mod~4). For the block structure of each class
	we refer to table~\ref{tb:RMTlist}. }
	\vspace{9pt}
	\renewcommand{\arraystretch}{1.05}
	\centering 
	\scalebox{0.92}{
	\begin{tabular}{|c||c|c|c|c|c|c|}
		\hline 
		\cellcolor{cgray}
		\!\!\!\! 
		$\begin{array}{c}\text{\footnotesize $\NN=1$ SYK} 
		\vspace{-4pt}\\ 
		{\footnotesize \qq\mequiv 1~\text{(mod~4)}} \end{array}$ \!\!\!\! 
		& $P^2$  
		& $R^2$  & $\begin{array}{c}\text{\footnotesize (anti-)}
		\vspace{-8pt}\\
		\text{\footnotesize \!\!\!\! commutators \!\!\!\!}\end{array}$
		& $\beta$ & {\footnotesize class of $Q$} &  
		$\begin{array}{c}\text{\footnotesize degeneracy}\vspace{-7pt}\\
		\text{\footnotesize \!\!\!\!of levels in $H$\!\!\!\!}\end{array}$ 
		\\\hline 
		\!$N_m\mequiv 0$ (mod~8)\! & $+1$ & $+1$ 
		& {\small $\begin{array}{c}\{P,Q\}=0\\{[R,Q]=0}\end{array}$} 
		& 1  & \!\!\!\!\!\!\!
		{\small $\begin{array}{c}\text{chGOE (BDI)}
		\vspace{-2pt}\\ \nu=0 \end{array}$} \!\!\!\!\!\!\!
		& 2
		\\\hline 
		\!$N_m\mequiv 2$ (mod~8)\! & $+1$ & $-1$ 
		& {\small $\begin{array}{c}{[P,Q]=0} \\ \{R,Q\}=0 \end{array}$}  
		& 1 & {\small BdG (CI)} & 2 
		\\\hline
		\!$N_m\mequiv 4$ (mod~8)\! & $-1$ & $-1$ 
		& {\small $\begin{array}{c}\{P,Q\}=0\\{[R,Q]=0}\end{array}$} 
		& 4 & {\small $\begin{array}{c}\text{chGSE (CII)}
		\vspace{-2pt}\\ \nu=0 \end{array}$} 
		& 4 
		\\\hline 
		\!$N_m\mequiv 6$ (mod~8)\! & $-1$ & $+1$ 
		& {\small $\begin{array}{c}{[P,Q]=0} \\ \{R,Q\}=0 \end{array}$}   
		& 4 & {\small $\begin{array}{c}\text{BdG} 
		\vspace{-4pt}\\ \text{(DIII-even)}
		\end{array}$} & 4
		\\\hline
	\end{tabular}}
    \end{table}
    \begin{table}
	\caption{\label{tb:RMTQ01}
	Symmetry classification of $Q$ in the $\NN=1$ SYK model  
	for $\qq\mequiv 3$ (mod~4). This table is consistent with 
	\cite{Li:2017hdt}. For the block structure of each class
	we refer to table~\ref{tb:RMTlist}. }
	\vspace{9pt}
	\centering  
	\scalebox{0.92}{
	\begin{tabular}{|c||c|c|c|c|c|c|}
		\hline 
		\cellcolor{cgray}
		\!\!\!\! 
		$\begin{array}{c}\text{\footnotesize $\NN=1$ SYK} 
		\vspace{-4pt}\\ 
		{\footnotesize \qq\mequiv 3~\text{(mod~4)}} \end{array}$ \!\!\!\! 
		& $P^2$  
		& $R^2$  & $\begin{array}{c}\text{\footnotesize (anti-)}
		\vspace{-8pt}\\
		\text{\footnotesize \!\!\!\! commutators \!\!\!\!}\end{array}$
		& $\beta$ & {\footnotesize class of $Q$} &  
		$\begin{array}{c}\text{\footnotesize degeneracy}\vspace{-7pt}\\
		\text{\footnotesize \!\!\!\!of levels in $H$\!\!\!\!}\end{array}$ 
		\\\hline 
		\!$N_m\mequiv 0$ (mod~8)\! & $+1$ & $+1$ 
		& {\small $\begin{array}{c}[P,Q]=0\\\{R,Q\}=0\end{array}$} 
		& 1  & {\small $\begin{array}{c}\!\!\!\text{chGOE (BDI)}\!\!\!
		\vspace{-2pt}\\ \nu=0 \end{array}$} 
		& 2
		\\\hline 
		\!$N_m\mequiv 2$ (mod~8)\! & $+1$ & $-1$ 
		& {\small $\begin{array}{c}\{P,Q\}=0 \\ {[R,Q]=0}\end{array}$}  
		& 4 & {\small $\begin{array}{c}\text{BdG}
		\vspace{-4pt}\\\text{(DIII-even)}\end{array}$} 
		& 4 
		\\\hline
		\!$N_m\mequiv 4$ (mod~8)\! & $-1$ & $-1$ 
		& {\small $\begin{array}{c}[P,Q]=0\\\{R,Q\}=0\end{array}$} 
		& 4 & {\small $\begin{array}{c}\text{chGSE (CII)}
		\vspace{-2pt}\\ \nu=0 \end{array}$} 
		& 4 
		\\\hline 
		\!$N_m\mequiv 6$ (mod~8)\! & $-1$ & $+1$ 
		& {\small $\begin{array}{c}\{P,Q\}=0 \\ {[R,Q]=0}\end{array}$}   
		& 1 & {\small BdG (CI)} 
		& 2
		\\\hline
	\end{tabular}}
\end{table}

One can also consider a superposition of multiple fermionic operators 
in the supercharge, e.g, $Q= i \sum_{ijk} C_{ijk}\chi_i \chi_j \chi_k 
+ \sum_{i}D_{i}\chi_i$, where $\{C_{ijk}\}$ and $\{D_i\}$ are 
independent real Gaussian couplings. Then $Q$ fails to 
commute or anti-commute with $P$ and $R$ and the symmetry class 
is changed: $Q$ now belongs to the $\beta=2$ 
chGUE (AIII) class with $\nu=0$. There is no degeneracy of eigenvalues 
for $Q$ while all eigenvalues of $H=Q^2$ are two-fold degenerate 
since $\{(-1)^F, Q\}=0$. 

In all cases considered above for $\NN=1$, the symmetry classes differ 
from the Wigner-Dyson classes because of the presence of 
chiral symmetry $(-1)^F$. This difference manifests itself 
in distinctive level correlations near the origin (universality at the hard edge). 
In order to expose this in the thermal $\NN=1$ SYK model, the temperature 
must be lowered to the scale of the smallest eigenvalue of $H$. This is 
exponentially small in $N_m$.

\subsection{Numerical simulations}

\begin{figure}[tbh]
	\centering
	\includegraphics{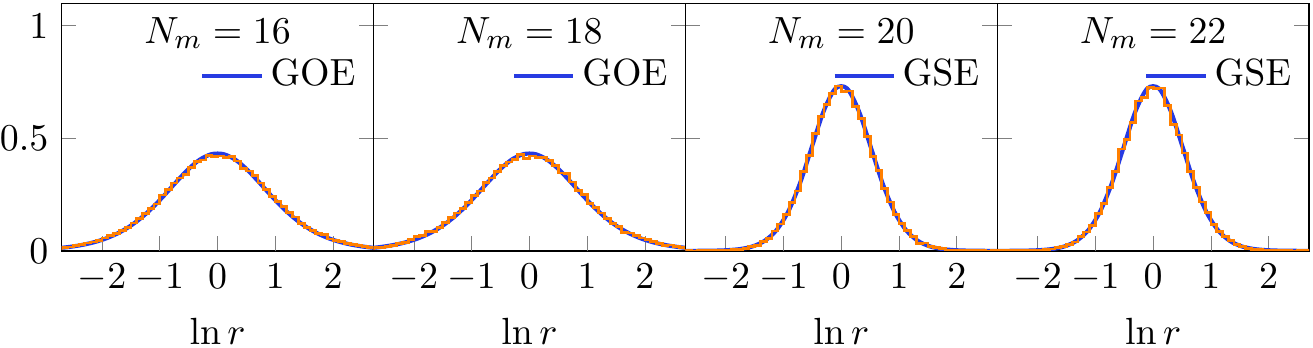}
	\vspace{-5pt}
	\caption{\label{fg:Plnr_N1}Distribution of the ratio $r$ of 
	two consecutive level spacings in the $\NN=1$ SYK model with $\qq=5$. 
	The number of realizations used for averaging was $10^3$ for $N_m=16$, 
	$100$ for $N_m=18$ and $22$, and $200$ for $N_m=20$. 
	The blue lines are surmises for the RMT classes in table~\ref{tb:RMTQ00}.}
\end{figure}

\paragraph{Level correlations in the bulk} \ \\
Previously, the level statistics in the bulk of the energy spectrum for the $\NN=1$ 
SYK model with $\qq=3$ was studied in \cite{Li:2017hdt} 
and results consistent with table~\ref{tb:RMTQ01} were reported.  
Here we report the first numerical analysis of the bulk statistics for the $\NN=1$ SYK model 
with $\qq=5$ via exact diagonalization, to test table~\ref{tb:RMTQ00}. 
To identify the symmetry class, we again used 
the ratio of two consecutive level spacings. 
Our numerical results are displayed in figure~\ref{fg:Plnr_N1}. 
Excellent agreement with the RMT curves of the symmetry classes predicted by 
table~\ref{tb:RMTQ00} is observed. 
This evidences the existence of quantum chaotic dynamics in this model and corroborates 
our classification scheme.

\paragraph{Universality at the hard edge} \ \\
Next we proceed to the investigation of universality of the level distributions near the origin. 
In contrast to the $\NN=0$ SYK model, whose hard edge at $\E=0$ was in the middle of the spectrum, 
the fluctuations of the smallest eigenvalues of $Q$ (or $H$) are of direct physical significance  
for the low-temperature thermodynamics of the $\NN=1$ SYK model.  We have
\begin{figure}
	\centering 
        \scalebox{.96}{
        \begin{tabular}{r@{\qquad}r}
          \includegraphics{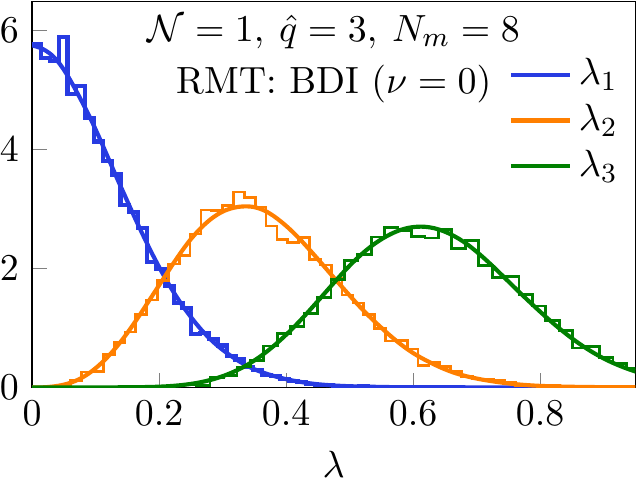} &
	  \includegraphics{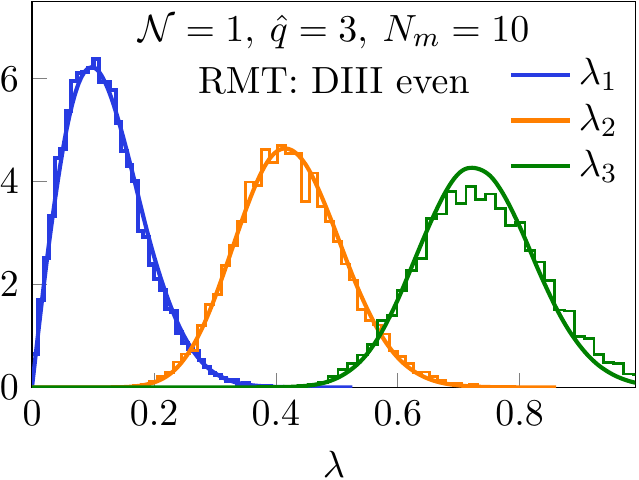} \\
	  \includegraphics{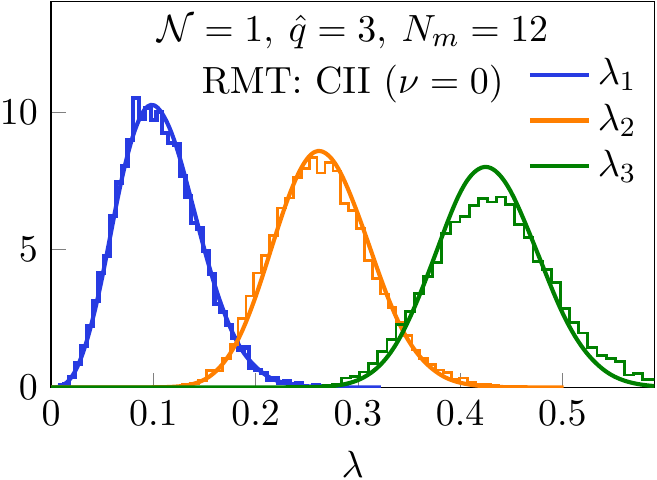} &
	  \includegraphics{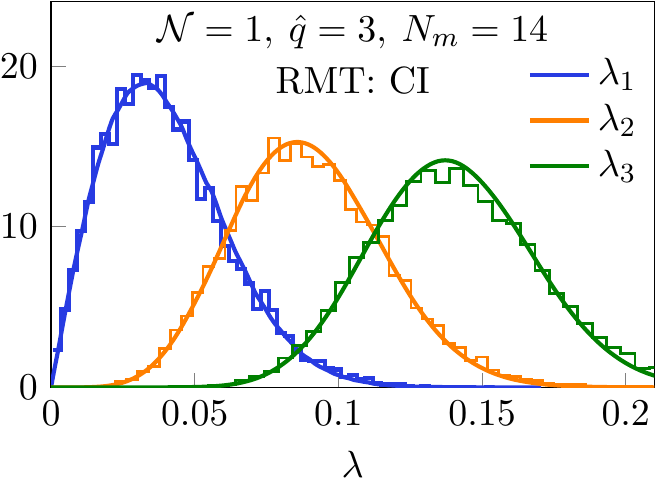}
        \end{tabular}}
	\vspace{-7pt}
	\caption{\label{fg:PminN1q3}Distributions of the smallest 3
          eigenvalues of $Q$ in \eqref{eq:QdefN1} in the $\NN=1$ SYK
          model with $\qq=3$ and $J=1$, compared with the predictions
          (solid lines) of the RMT classes in
          table~\ref{tb:RMTQ01}. The number of independent random
          samples is $10^4$ for each plot. As in
          figure~\ref{fg:PminN0}, the small deviations from RMT for
          $\lambda_3$ are interpreted to be effects of finite $N_m$.}
        \vspace*{1mm}
\end{figure}
\begin{figure}[H]
	\centering 
        \scalebox{.96}{
        \begin{tabular}{r@{\qquad}r}
	\includegraphics{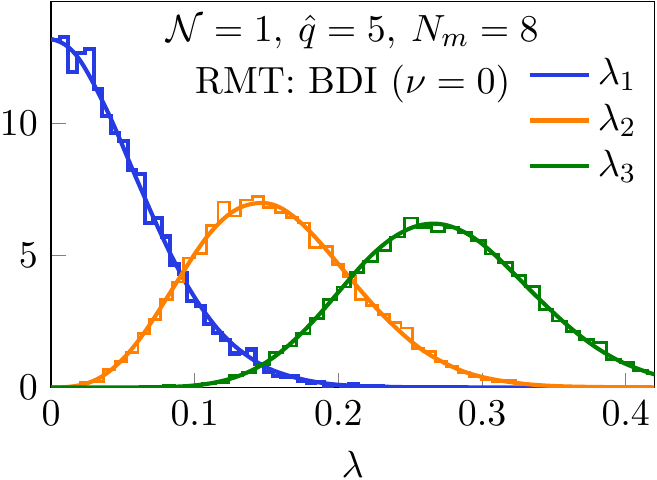} &
	\includegraphics{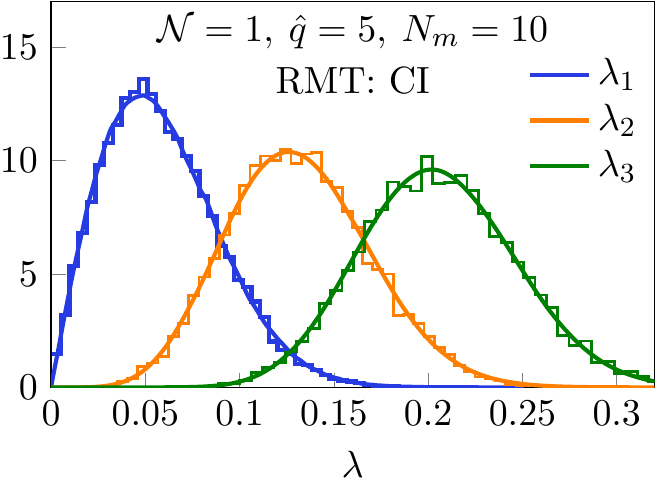} \\
	\includegraphics{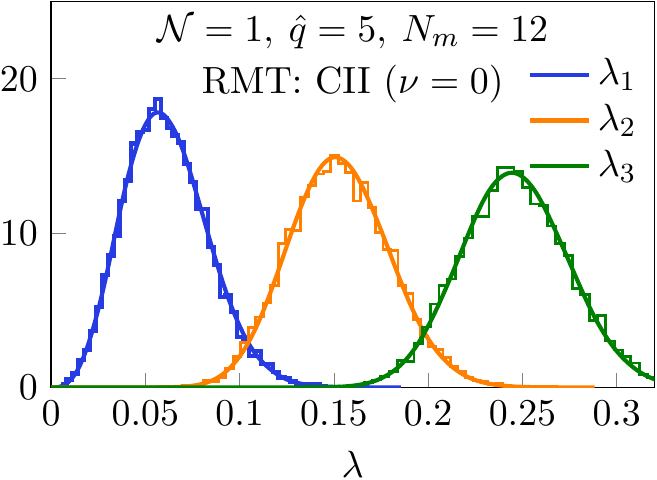} & 
	\includegraphics{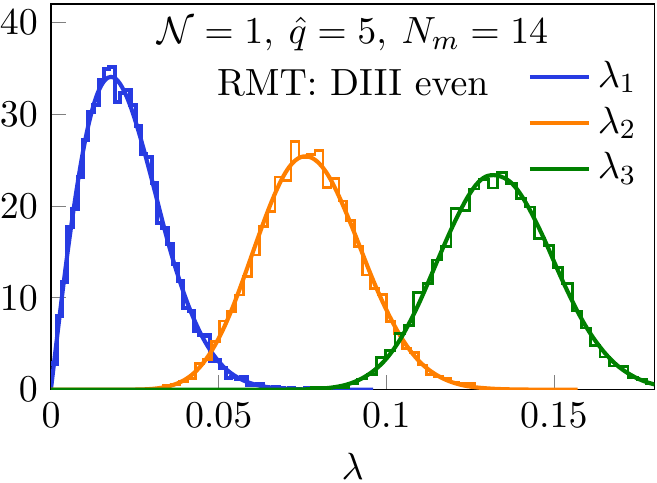}
        \end{tabular}}
	\vspace{-7pt}
	\caption{\label{fg:PminN1q5}Same as figure~\ref{fg:PminN1q3}
          but for $\qq=5$ and compared with the RMT predictions in
          table~\ref{tb:RMTQ00}.}
\end{figure}
\noindent 
numerically studied the distributions of the smallest three eigenvalues 
of $Q$ for the $\NN=1$ SYK model with $\qq=3$ and $5$ for varying $N_m$.  
(The twofold degeneracy of each level was resolved in the case of $\beta=4$.) 
The results for $\qq=3$ and $5$ are shown in figures~\ref{fg:PminN1q3} and 
\ref{fg:PminN1q5}, respectively. They show very good agreement 
with the corresponding RMT predictions in tables~\ref{tb:RMTQ01} and \ref{tb:RMTQ00}.
The smallest eigenvalue approaches zero from above for larger $N_m$, 
indicating restoration of SUSY in the large-$N_m$ limit as already reported in \cite{Fu:2016vas}.  

We note that the RMT classes chGOE (BDI) and chGSE (CII) were originally invented  
and exploited in attempts to theoretically understand fluctuations of small eigenvalues 
of the Euclidean QCD Dirac operator with special antiunitary symmetries in a finite volume 
\cite{Verbaarschot:1994qf,Verbaarschot:1994ia,Halasz:1995qb,Verbaarschot:1997bf,
Nagao:2000cb},%
\footnote{See also \cite{Nagao:1991xj,NagaoSlevin1993,Forrester:1993vtx,
Nagao:1995np,Nagao:1998fbu} 
for related works in mathematics.} 
related to spontaneous breaking of chiral symmetry through the Banks-Casher 
relation \cite{Banks:1979yr}. 
The RMT predictions agree well with the Dirac spectra taken from lattice QCD 
simulations \cite{BerbenniBitsch:1997tx}. It is a nontrivial observation
that the smallest energy levels of the $\NN=1$ SYK model, 
which set the scale for the spontaneous breaking of SUSY, obey the same statistics 
as the eigenvalues of the Dirac operator in QCD, which has totally different microscopic 
interactions compared to the SYK model. This is yet another example for 
random matrix universality. 

\section{\boldmath Interlude: a simple model bridging the gap between $\NN=1$ and $2$\label{sc:interlude}}
\subsection{Motivation and definition}

The SYK model with $\NN=2$ SUSY \cite{Fu:2016vas} has the Hamiltonian  
$H=\{Q,\bar{Q}\}$ with two supercharges $Q$ and $\bar{Q}$, each 
comprising an odd number of \emph{complex} fermions. This model preserves 
the $\U(1)$ fermion number exactly, so that the Hamiltonian is block-diagonal 
in the fermion-number eigenbasis. As shown by the Witten-index computation in \cite{Fu:2016vas}, 
the Hamiltonian has an extensive number of exact zero modes%
\footnote{The existence of a macroscopic number of ground states is 
a familiar phenomenon in lattice models with exact SUSY 
\cite{Nicolai:1976xp,Nicolai:1977qx,Fendley:2002sg,Fendley:2003je,Fendley:2005ae,Eerten2005}.} 
and SUSY is unbroken at finite $N_c$\,. 
These features are in marked contrast to the $\NN=1$ SYK model, 
where the fermion number is only conserved modulo 2, the Hamiltonian is positive definite 
with no exact zero modes, and SUSY is spontaneously broken at finite $N_c$\,. 

While there is no logical obstacle to moving from $\NN=1$ to $2$, it is
helpful to have a simple model that serves as a bridge between 
these two theories. The model we designed for this purpose is defined 
by the Hamiltonian $H=M^2$ with the Hermitian operator 
\ba
	M \equiv i^{p/2} \!\!\!\!\! \sum_{1\leq j_1<\cdots <j_p \leq N_c}
	\!\!\!\! \big( 
		Z_{j_1\cdots j_p}\, c_{j_1} \cdots c_{j_p} 
		+ \bar{Z_{j_1\cdots j_p}}\,\bar{c}_{j_1} \cdots \bar{c}_{j_p}
	\big)\,,
	\label{eq:Mdefin}
\ea
where $1\leq p \leq N_c$ is an \emph{even} integer and 
$Z_{j_1\cdots j_p}$ are independent complex Gaussian random variables with 
mean zero and $\langle \bar{Z_{ab}}Z_{ab} \rangle=2J/N_c^2$ for some $J>0$. 
The creation and annihilation operators $\bar{c}_a$ and $c_a$ were introduced
in section~\ref{sc:PHT}. Because of $M=M^\dagger$ we have
$H\geq 0$, similarly to the supersymmetric SYK models. 
If we forcefully substitute $p=3$ and let $i^{p/2}\to i$, then 
$M=Q+\bar{Q}$ and $H=M^2=\{Q,\bar{Q}\}$, 
i.e., the $\NN=2$ SYK model is recovered (see section~\ref{sc:N2SYK}). 
What difference emerges if we retain an even number of fermions in $M$? 
Of course it makes $M$ a bosonic operator and destroys SUSY. 
At this cost, however, we gain three new features that were missing 
in the $\NN=1$ SYK model: (i) the fermion number 
is conserved modulo $2p$ (rather than modulo $2$), 
(ii) $H$ has a large number of \emph{exact} zero modes, and 
(iii) an interplay between $N_c$ and $F$ emerges in the 
symmetry classification of energy-level statistics. The last point is especially 
intriguing since this property is shared by the $\NN=2$ SYK model 
(section~\ref{sc:N2SYK}). This is why we 
regard this model as ``intermediate'' between the $\NN=1$ and $\NN=2$ SYK models. 
Studying the level structure of this exotic model provides a useful digression before 
tackling the $\NN=2$ case. 

\begin{figure}[tb]
	\centering
        \begin{tabular}{r@{\hspace*{15mm}}r}
	\includegraphics[height=44mm]{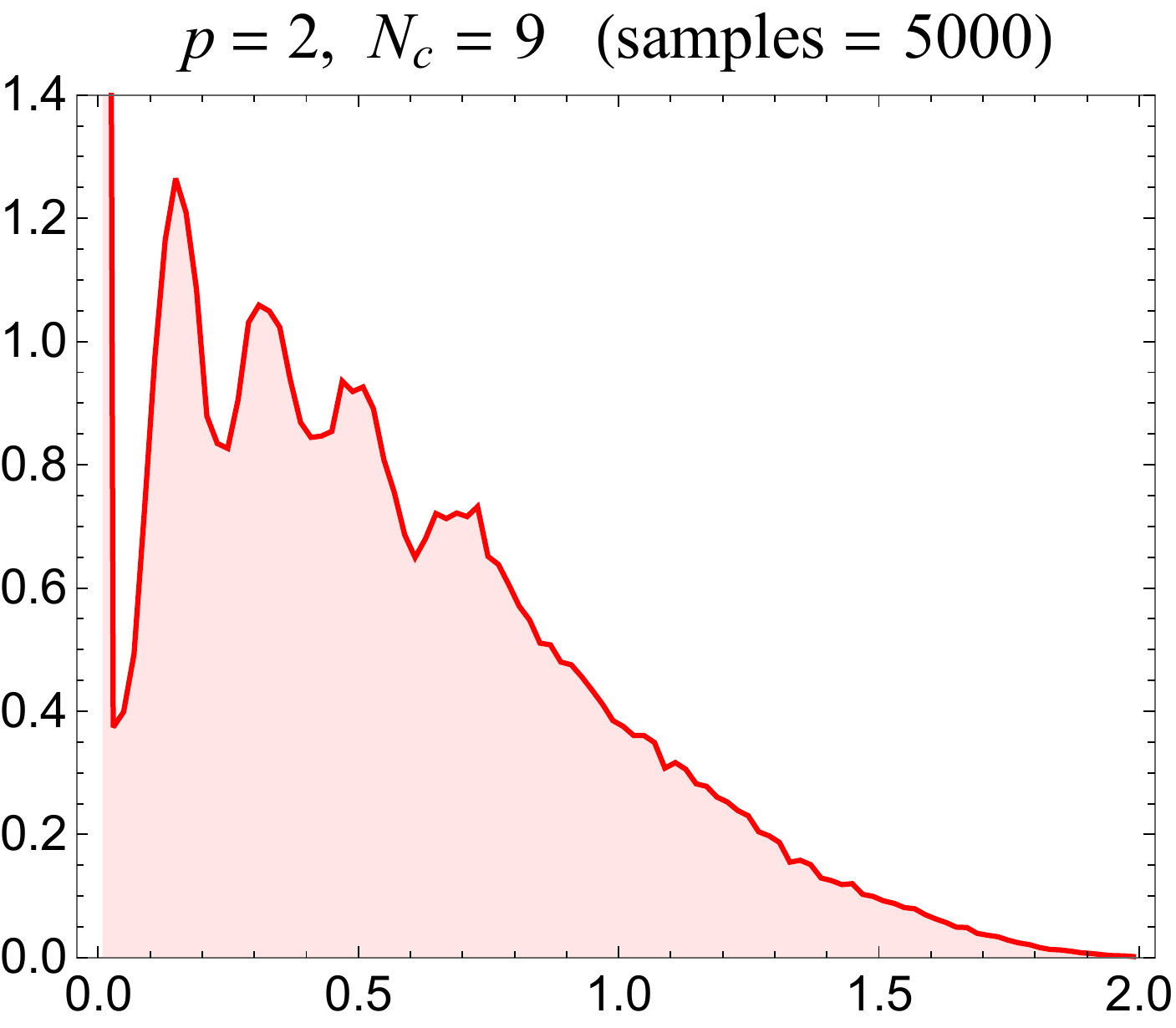} &
	\includegraphics[height=44mm]{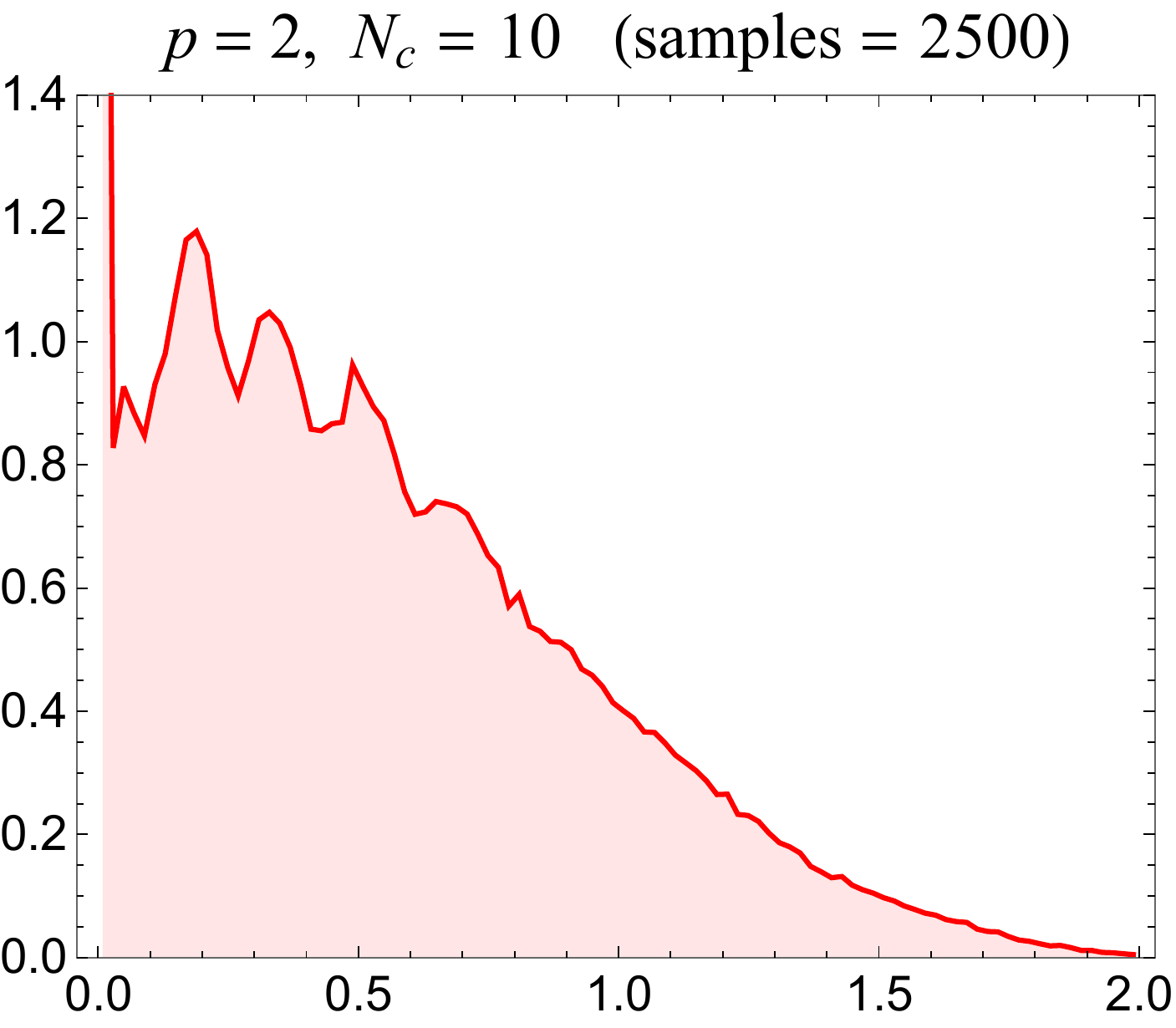} \\
	\includegraphics[height=44mm]{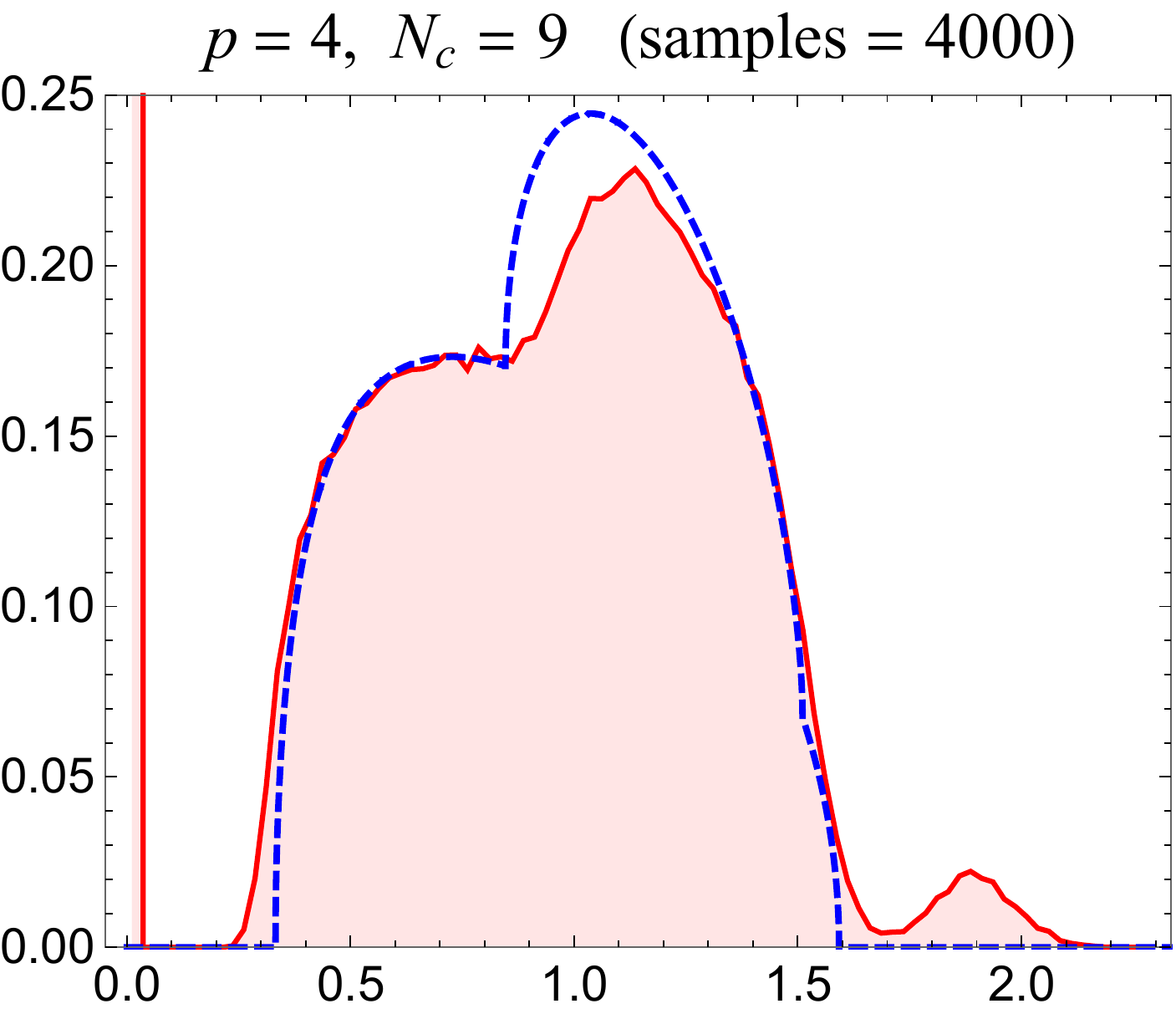}\hspace*{2pt} &
	\includegraphics[height=44mm]{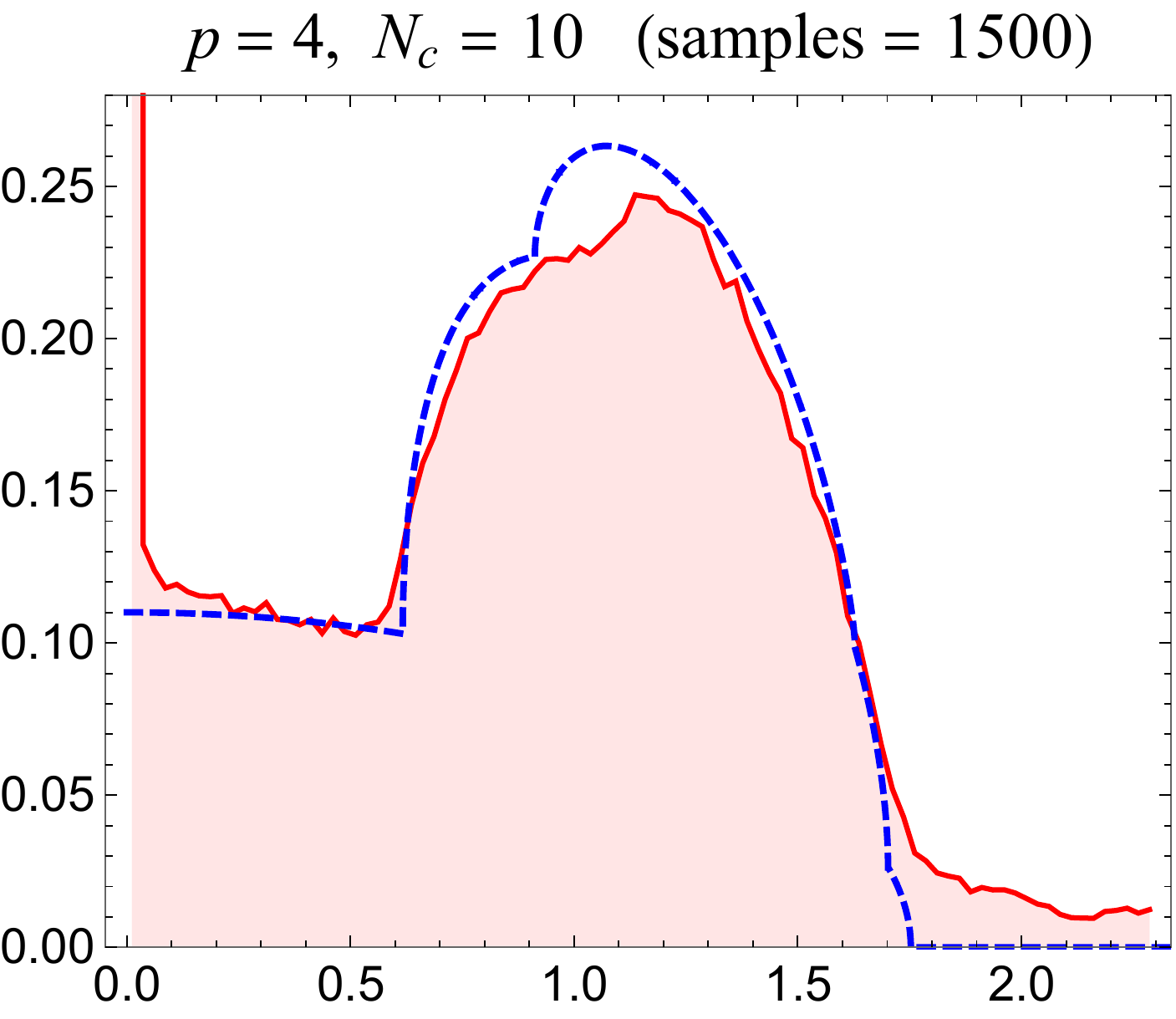}\hspace*{2pt}
        \end{tabular}
	\caption{\label{fg:totaldensity910}Spectral density of $M$ in \eqref{eq:Mdefin}
	for $p=2$ (top) and $4$ (bottom) at $N_c=9$ and $10$, averaged over many random samples. 
	Since the spectra are symmetric about $0$, only the nonnegative part is shown.  
	The sharp peak at the origin in each plot represents the density of 
	exact zero modes. In all plots $J=1$ and the total density is normalized to 1. 
	The blue dashed lines in the bottom plots are analytic approximations \eqref{eq:Pp4appro} 
	based on the Mar\v{c}enko-Pastur law.}
\end{figure}

By exact diagonalization we have numerically computed 
the spectral density of $M$ for $p=2$ and $4$, see figure~\ref{fg:totaldensity910}. 
In all plots there is a delta function at zero due to the macroscopic number of 
zero-energy states. Interestingly, the global shape never resembles  
Wigner's semicircle but rather depends sensitively on both $p$ and $N_c$. 
For $p=2$ we observe oscillations in the middle of 
the spectrum, for which we currently do not have a simple explanation. 
The case $p=2$ could be more the exception than the rule,%
\footnote{We speculate that the spectral density for this case may even 
be computed exactly since $M$ is just a fermion bilinear, 
but this is beyond the scope of this paper.} much like the $q=2$ SYK model 
that is solvable and nonchaotic \cite{Maldacena:2016hyu,
Gross:2016kjj,Cotler:2016fpe} unlike its $q>2$ counterparts.  

For both $p=2$ and $4$, a close inspection of the plots near the origin 
reveals that for odd $N_c$ there is a dip of the density around the origin, indicating 
that small nonzero levels are \emph{repelled from the origin}, while 
there is no such repulsion for even $N_c$\,. The same tendency of 
the spectral density (albeit with the parity of $N_c$ reversed) has been observed 
for the $\NN=2$ SYK model, too \cite{Stanford:2017thb}. We will give 
a simple explanation of this phenomenon later.

\subsection[Classification for $p=2$]{\boldmath Classification for $p=2$}

To make the presentation as simple as possible, we shall begin with 
$p=2$, in which case the fermion number $F$ is conserved modulo $4$. 
The Hilbert space $V$ of $N_c$ complex fermions can be arranged into 
a direct sum of four spaces $\VV_{0,1,2,3}$, where $\VV_f$ is the eigenspace of $F$  
corresponding to $F = f$ (mod~4), i.e., 
\ba
	V=\VV_0\oplus \VV_1 \oplus \VV_2 \oplus \VV_3 
\ea
with $\dim(V) = \sum_{f=0}^{3}\dim(\VV_f) = 2^{N_c}$ and
\begin{align}\label{eq:dimsp2}
  \DD_f \equiv \mathrm{dim}(\VV_f) = \!\!\! 
  \sum_{k=0}^{\lfloor (N_c-f)/4 \rfloor}\binom{N_c}{4k+f}\,.
\end{align}
The numbers $\DD_{0,1,2,3}$ are listed for $3\leq N_c\leq 10$ in table~\ref{tb:toymodelp2}. 
Since there is no nonzero matrix element of $M$ between states with different parity of $F$
we have $M=\scalebox{0.8}{$\displaystyle 
\bep \0 & A_0 \\ A_0^\dagger & \0 \eep \oplus 
\bep \0 & A_1 \\ A_1^\dagger & \0 \eep$}$, where the first (second) term corresponds 
to $\VV_0\oplus \VV_2$ ($\VV_1\oplus \VV_3$). 
The chiral structure in each term is due to the chiral symmetry $\{i^F,M\}=0$,  
which ensures the spectral mirror symmetry of $M$.

It should be stressed that $A_0$ and $A_1$ are in general rectangular. 
When they become a square matrix can be 
read off from table~\ref{tb:toymodelp2}. These cases are colored in red and green. 
They only occur for even $N_c$ (which is also true for $p=4$,
see table~\ref{tb:toymodelp4} below). 
On the other hand, for odd $N_c$\,, both $A_0$ and $A_1$ are rectangular. 
As is well known from studies in chiral RMT \cite{Verbaarschot:1997bf,Verbaarschot:2000dy},  
in that case the nonzero eigenvalues of $M$ (i.e., the nonzero singular 
values of $A_0$ and $A_1$) are pushed away from the origin 
by the large number of exact zero modes. Indeed, $\alpha$ in table~\ref{tb:RMTlist} is 
proportional to the number of zero modes, and large $\alpha$ suppresses the joint 
probability density of eigenvalues near zero. 
This leads to the dip around the origin in the left plots of figure~\ref{fg:totaldensity910}. 
However, for even $N_c$\,, 
in the subspaces without exact zero modes there is no repulsion of the nonzero modes from the origin, 
and thus no dip of the density (which is summed over all subspaces) shows up near zero. 

In order to understand the level degeneracy in each sector correctly, 
we must figure out the antiunitary symmetries of the matrix $M$. 
We use the particle-hole operator $P$ in \eqref{eq:defP} again. In addition, 
we define another antiunitary operator $S\equiv P\cdot i^F$.  One can show
\ba
	\{P,M\}=0 \qquad \text{and} \qquad [S,M]=0 
	\qquad \text{for all $N_c$\,.}
\ea
Both $P^2$ and $S^2$ are tabulated in table~\ref{tb:toymodelp2}, but 
extra care is needed for $S$ because $S^2$ is not just $\pm 1$ but 
a nontrivial operator that depends on $F$. 
\begin{table}[bt]
	\caption{\label{tb:toymodelp2}Model \eqref{eq:Mdefin} for $p=2$. 
	We list the dimensions \eqref{eq:dimsp2} of the eigenspaces of $F$ (mod~4). 
	Uncolored blocks belong to chGUE\,(AIII)$_{\beta=2}$, while 
	\raisebox{2pt}{\colorbox[gray]{0.85}{\;\;\;\tiny$\strut$}}:~chGSE\,(CII)$_{\beta=4}$ 
	with $\nu=|\DD_1-\DD_3|/2$, 
	\raisebox{2pt}{\colorbox[rgb]{0.8,1,0.8}{\;\;\;\tiny$\strut$}}:~BdG\,(DIII-even)$_{\beta=4}$,
	\raisebox{2pt}{\colorbox[rgb]{0.7,0.9,1}{\;\;\;\tiny$\strut$}}:~chGOE\,(BDI)$_{\beta=1}$ 
	with $\nu=|\DD_0-\DD_2|$, and
	\raisebox{2pt}{\colorbox[rgb]{1,0.8,0.8}{\;\;\;\tiny$\strut$}}:~BdG\,(CI)$_{\beta=1}$. 
	Details of each class can be found in table~\ref{tb:RMTlist}.
        Also shown are the squares of the antiunitary operators $P$ and $S$.
	The symmetry pattern is periodic in $N_c$ with period $4$.}
	\vspace{10pt}
	\centering
	\scalebox{0.9}{
	\renewcommand{\arraystretch}{1.1}
	\begin{tabular}{|c|cccccccc|}
          \hline
		$N_c$ &3&4&5&6&7&8&9&10 
		\\\hhline{|=|========|}
		$\DD_0$ &1&\cellcolor[rgb]{0.7,0.9,1}2&6&\cellcolor[rgb]{1,0.8,0.8}16&36
		&\cellcolor[rgb]{0.7,0.9,1}72
		&136& \cellcolor[rgb]{1,0.8,0.8}256
		\vspace{-2pt}
		\\
		$\DD_2$ &3&\cellcolor[rgb]{0.7,0.9,1}6&10&\cellcolor[rgb]{1,0.8,0.8}16&28
		&\cellcolor[rgb]{0.7,0.9,1}56
		&120& \cellcolor[rgb]{1,0.8,0.8}256
		\\\hline 
		\!\!\!{\footnotesize $\#$Zero modes}\!\!\! &2&4&4&0&8&16&16&0
		\\\hhline{|=|========|}
		$\DD_1$ &3& \cellcolor[rgb]{0.8,1,0.8}4&6
		&\cellcolor[gray]{0.85}12&28&
		\cellcolor[rgb]{0.8,1,0.8}64&136& \cellcolor[gray]{0.85}272
		\vspace{-2pt}
		\\
		$\DD_3$ &1& \cellcolor[rgb]{0.8,1,0.8}4&10&\cellcolor[gray]{0.85}20&36&
		\cellcolor[rgb]{0.8,1,0.8}64&120& \cellcolor[gray]{0.85}240
		\\\hline 
		\!\!\!{\footnotesize $\#$Zero modes}\!\!\! &2&0&4&8&8&0&16&32
		\\\hhline{|=|========|}
		$P^2$ & $-1$ & 1 & 1 & $-1$ &$-1$&1&1&$-1$
		\vspace{-2pt}
		\\
		$S^2$ & $(-1)^{F+1}i$ & $(-1)^{F}$
		& $(-1)^{F+1}i$
		& $(-1)^{F}$ & $(-1)^{F+1}i$ 
		& $(-1)^{F}$ & $(-1)^{F+1}i$ & $(-1)^{F}$
		\\\hline 
	\end{tabular}}
\end{table}

For even $N_c$\,, each chiral block belongs to one of  
chGSE\,(CII)$_{\beta=4}$, BdG\,(DIII-even)$_{\beta=4}$, 
chGOE\,(BDI)$_{\beta=1}$, and BdG\,(CI)$_{\beta=1}$ according to 
the values of $P^2$ and $S^2$ (cf.~table~\ref{tb:RMTlist}).  
In the $\beta=4$ classes, every nonzero level must come in 
quadruplets $(\lambda,\lambda,-\lambda,-\lambda)$ due to 
Kramers degeneracy and chiral symmetry. 

For odd $N_c$\,, $P$ maps a state in $\VV_0\oplus \VV_2$ 
to $\VV_1\oplus \VV_3$ and vice versa. Therefore the nonzero levels 
of $M$ in $\VV_0\oplus \VV_2$ must be degenerate with those in 
$\VV_1 \oplus \VV_3$.  Since there is no antiunitary symmetry acting within each 
chiral block, all uncolored sectors in table~\ref{tb:toymodelp2} 
belong to chGUE\,(AIII). 

This completes the algebraic classification of the model 
\eqref{eq:Mdefin} for $p=2$ based on RMT. This classification 
is periodic in $N_c$ with period $4$ as can be seen from 
table~\ref{tb:toymodelp2}. We have numerically checked the level degeneracy 
of $M$ in each sector for various $N_c$ and confirmed consistency with our 
classification. In this process we found, surprisingly, that 
levels often show a large (e.g., 16-fold) degeneracy that cannot be 
accounted for by our antiunitary symmetries $P$ and $S$. Such a large 
degeneracy, which presumably is responsible for the wavy shape in 
the upper plots of figure~\ref{fg:totaldensity910} and makes the level 
spacing distribution for $p=2$ deviate from RMT, was not observed for $p=4$. 
We interpret this as an indication that the model with $p=2$ is just too simple 
to show quantum chaos and therefore do not investigate it further. 

\subsection[Classification for $p=4$]{\boldmath Classification for $p=4$}

As a more nontrivial case we now study the $p=4$ model, which 
preserves $F$ (mod~8). This time the Hilbert space decomposes as  
$V=\bigoplus\limits_{f=0}^7\VV_f$ with 
\begin{align}\label{eq:dimsp4}
  \DD_f \equiv \dim(\VV_f)=
  \sum_{k=0}^{\lfloor (N_c - f)/8 \rfloor}\!\!\binom{N_c}{8k+f}\,.
\end{align}
$M$ acquires a block-diagonal form, $M=\scalebox{0.8}{$\displaystyle 
\bep\0& A_0 \\ A_0^\dagger & \0\eep\oplus 
\bep\0& A_1 \\ A_1^\dagger & \0\eep\oplus 
\bep\0& A_2 \\ A_2^\dagger & \0\eep\oplus 
\bep\0& A_3 \\ A_3^\dagger & \0\eep$}$, where the terms correspond to 
$\VV_0\oplus \VV_4$, $\VV_1\oplus \VV_5$, 
$\VV_2\oplus \VV_6$, and $\VV_3\oplus \VV_7$, respectively. As a consequence,
the spectrum of $M$ enjoys a mirror symmetry as in the 
model with $p=2$. Let us define an antiunitary operator 
$\wt{S}\equiv P\cdot \kappa^F$, where 
$\kappa\equiv\rme^{i \pi/4}$ is the 8-th root of unity and 
$P$ was defined in \eqref{eq:defP}.  One can show
\ba
	[P,M]=0 \qquad \text{and} \qquad \{ \wt{S},M \}=0 
	\qquad \text{for all $N_c$\,.}
\ea

\begin{table}[tb]
	\caption{\label{tb:toymodelp4}Model \eqref{eq:Mdefin} for $p=4$. 
	We list the dimensions \eqref{eq:dimsp4} of the eigenspaces of $F$ (mod~8). 
	Uncolored blocks belong to chGUE\,(AIII)$_{\beta=2}$, while 
	\raisebox{2pt}{\colorbox[gray]{0.85}{\;\;\;\tiny$\strut$}}:~chGSE\,(CII)$_{\beta=4}$ 
	with $\nu=|\DD_i - \DD_{i+4}|/2$, 
	\raisebox{2pt}{\colorbox[rgb]{0.8,1,0.8}{\;\;\;\tiny$\strut$}}:~BdG\,(DIII-even)$_{\beta=4}$,
	\raisebox{2pt}{\colorbox[rgb]{0.7,0.9,1}{\;\;\;\tiny$\strut$}}:~chGOE\,(BDI)$_{\beta=1}$ 
	with $\nu=|\DD_i - \DD_{i+4}|$, and
	\raisebox{2pt}{\colorbox[rgb]{1,0.8,0.8}{\;\;\;\tiny$\strut$}}:~BdG\,(CI)$_{\beta=1}$. 
	Details of each class can be found in table~\ref{tb:RMTlist}.
	The mark $(\alert{2})$ after the number of positive levels of $M$ indicates that those levels are twofold degenerate, e.g., 
	$20\,(\alert{2})$ means $10$ pairs. In each block of given $N_c$ 
	there is an equal number of positive and negative levels because of chiral symmetry, 
	$\{\kappa^F, M\}=0$.
        Also shown are the squares of the antiunitary operators $P$ and $\wt{S}$.
	The symmetry pattern is periodic in $N_c$ with period $8$.}
	\vspace{5pt}
	\centering 
	\scalebox{0.9}{\noindent
	\renewcommand{\arraystretch}{1.1}
	\begin{tabular}	{|c|cccccccc|}
          \hline
		$N_c$ &7&8&9&10&11&12&13 & 14%
		\\ \hhline{|=|========|}
		$\DD_0$ &1&\cellcolor[rgb]{0.7,0.9,1}2&10&46&166
		& \cellcolor[rgb]{1,0.8,0.8}496&1288 & 3004%
		\vspace{-2pt}
		\\
		$\DD_4$ &35&\cellcolor[rgb]{0.7,0.9,1}70&126&210&330
		&\cellcolor[rgb]{1,0.8,0.8}496&728 & 1092%
		\\ \hline 
		\!\!\scalebox{0.7}{$\begin{array}{c}\text{$\#$\,Positive} 
		\vspace{-4pt}\\ \text{levels of $M$}\end{array}$}\!\!\!
		&1&2&10&46&166&496&728 & 1092%
		\vspace{-2pt} \\ \hhline{|=|========|}
		$\DD_1$ &7&8&10&\cellcolor[gray]{0.85}20&66&232&728 
		& \cellcolor[rgb]{0.8,1,0.8}2016%
		\vspace{-2pt}
		\\
		$\DD_5$ &21&56&126&\cellcolor[gray]{0.85}252&462&792&1288 
		& \cellcolor[rgb]{0.8,1,0.8}2016%
		\\ \hline 
		\!\!\scalebox{0.7}{$\begin{array}{c}\text{$\#$\,Positive} 
		\vspace{-4pt}\\ \text{levels of $M$}\end{array}$}\!\!\! 
		&7&8&10& \!\!\! 20\,(\alert{2})\!\!\! 
		&66&232&728 & \!\!2016\,(\alert{2})\!\!\!%
		\vspace{-2pt} \\ \hhline{|=|========|}
		$\DD_2$ &21& \! \cellcolor[rgb]{1,0.8,0.8}28 \! 
		&36&46&66&\cellcolor[rgb]{0.7,0.9,1}132&364 & 1092%
		\vspace{-2pt}
		\\
		$\DD_6$ &7&\cellcolor[rgb]{1,0.8,0.8}28
		&84&210&462&\cellcolor[rgb]{0.7,0.9,1}924&1716 & 3004%
		\\ \hline 
		\!\!\scalebox{0.7}{$\begin{array}{c}\text{$\#$\,Positive} 
		\vspace{-4pt}\\ \text{levels of $M$}\end{array}$}\!\!\!
		&7&28&36&46&66&132&364 & 1092%
		\vspace{-2pt} \\ \hhline{|=|========|}
		$\DD_3$ &35&56&84&\cellcolor[rgb]{0.8,1,0.8}120
		&166&232&364 & \cellcolor[gray]{0.85}728%
		\vspace{-2pt}
		\\
		$\DD_7$ &1&8&36&\cellcolor[rgb]{0.8,1,0.8}120
		&330&792&1716 & \cellcolor[gray]{0.85}3432%
		\\ \hline 
		\!\!\scalebox{0.7}{$\begin{array}{c}\text{$\#$\,Positive} 
		\vspace{-4pt}\\ \text{levels of $M$}\end{array}$}\!\!\!
		&1&8&36& \!\!\!120\,(\alert{2})\!\!\!
		&166&232&364 & \!\!728\,(\alert{2})\!\!%
		\vspace{-2pt} \\ \hhline{|=|========|}
		$P^2$ &$-1$&1&1&$-1$&$-1$&1&1&$-1$  
		\vspace{-2pt}
		\\
		$\wt{S}^2$ & \!\!$-\kappa^{2F+1}$\!\! 
		& \!$\kappa^{2F}$\! & $\kappa^{2F-1}$ & 
		\!\!\!\! $\kappa^{2F+2}$ \!\!\!\! & $\kappa^{2F+1}$ & 
		\!\!$-\kappa^{2F}$\!\! & $\kappa^{2F+3}$ & $\kappa^{2F-2}$  
		\\ \hline 
	\end{tabular}}
\end{table}

The dimension of each subspace of $V$ is listed for $7\leq N_c \leq 14$ 
in table~\ref{tb:toymodelp4}. As for $p=2$, 
the particle-hole operator $P$ generates degeneracies between distinct chiral 
blocks. For instance, at $N_c=11$, the 166 distinct positive levels in $\VV_0\oplus \VV_4$ 
are degenerate with those in $\VV_3\oplus \VV_7$. The symmetry classification is 
just a rerun of our arguments for $p=2$ and therefore omitted here. We have 
numerically confirmed that table~\ref{tb:toymodelp4} gives the 
correct degeneracy of levels. (Unlike for $p=2$, we did not observe any 
unexpected further degeneracies.) 

\subsection{Global spectral density}

Table~\ref{tb:toymodelp4} not only provides a symmetry classification but also 
enables us to derive a fairly simple analytic approximation 
to the global spectral density. Let us recall the so-called 
Mar\v{c}enko-Pastur law \cite{MP1967}: suppose $X$ is a complex 
$L\times N$ matrix with $L\leq N$ whose elements are independently and 
identically distributed with $\langle X_{ij} \rangle=0$ and 
$\langle |X_{ij}|^2 \rangle=\sigma^2<\infty$. Let us denote the $L$ 
eigenvalues of $\sqrt{XX^\dagger}$ by $\{\xi_{i}\}\geq 0$. Then for
$L,N\to\infty$ with $L/N\in(0,1]$ fixed, 
the probability distribution of $\{\xi_i\}$ takes on the limit
\ba
	P_{L,N}(\sigma;\xi) = \frac{1}{\sqrt{N}\sigma}F \mkakko{\frac{L}{N}, \frac{\xi}{\sqrt{N}\sigma}},
	\label{eq:Pmpl}
\ea
where
\ba
	F(\alpha,x) \equiv \begin{cases}
		\displaystyle 
		\frac{1}{\pi \alpha x}\sqrt{\kkakko{(1+\sqrt{\alpha})^2-x^2}\kkakko{x^2-(1-\sqrt{\alpha})^2}}
		& \text{for } x\in [1-\sqrt{\alpha}, 1+\sqrt{\alpha}\,]\,,
		\\ ~~0 & \text{otherwise}\,.
	\end{cases}
\ea
This function satisfies the normalization $\int_0^\infty \rmd x\, F(\alpha,x)=1$ for all $\alpha\in(0,1]$. 
We now exploit this law to describe the global density of our $p=4$ model, 
shown previously in figure~\ref{fg:totaldensity910}. Whether \eqref{eq:Pmpl} works quantitatively or not is 
not obvious a priori because the matrix elements of \eqref{eq:Mdefin} are far from 
statistically independent, but rather strongly correlated. 
Putting this worry aside, let us consider the $N_c=9$ case first. 
According to table~\ref{tb:toymodelp4}, there are four chiral blocks, and two of them  
are copies of the other two, so we should sum just two Mar\v{c}enko-Pastur 
distributions. For $N_c=10$, we have to sum three.   
Taking into account that the global density in figure~\ref{fg:totaldensity910} 
counts both positive modes and exact zero modes, we obtain formulas with the correct 
normalization,
\begin{subequations}
\label{eq:Pp4appro}
\ba
	P^{(p=4,N_c=9)}(\sigma; \xi) & = \frac{2 \big[
		10\cdot P_{10,126}(\sigma;\xi) + 36 \cdot P_{36,84}(\sigma;\xi) 
	\big]}{2^9-2(10+36)}\,,
	\\
	P^{(p=4,N_c=10)}(\sigma; \xi) & = \frac{
		2\cdot 46 \cdot P_{46,210}(\sigma;\xi) + 20 \cdot P_{20,252}(\sigma;\xi) 
		+ 120 \cdot P_{120,120}(\sigma;\xi) 
	}{2^{10}-(2\cdot 46+20+120)}\,.
\ea
\end{subequations}
The parameter $\sigma$ has to be tuned to achieve the best fit to the data 
because RMT does not know the typical energy scale of the model. The results 
of the fits displayed in the bottom plots of figure~\ref{fg:totaldensity910} 
show impressive quantitative agreement. We also notice a shortage of 
levels near the peak density, as well as a leakage of levels toward larger values. 
Even though the agreement is not perfect it is intriguing that a na\"ive ansatz 
such as \eqref{eq:Pp4appro} is sufficient to account for the shape of the global density. 
We tried a similar fit for $p=2$ as well but did not find any agreement 
even at a qualitative level, probably due  
to the nonchaotic character of the $p=2$ model as described before.  

\subsection{Numerical simulations}

\begin{figure}[tbh]
	\centering
	\includegraphics{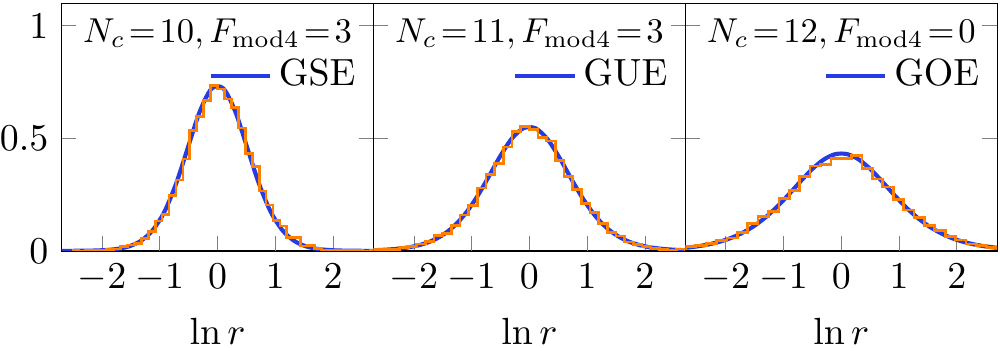}
	\vspace{-5pt}
	\caption{\label{fg:logrp4}Distribution of the ratio $r$ of 
	two consecutive level spacings of $M$ in \eqref{eq:Mdefin} with $p=4$. 
	The number of random samples used for averaging was 180 for $N_c=10$, 
	120 for $N_c=11$, and 40 for $N_c=12$. 
	The blue lines are surmises for the RMT classes 
	in table~\ref{tb:toymodelp4}.}
\end{figure}

\paragraph{Level correlations in the bulk} \ \\
We numerically checked the bulk statistics (GOE/GUE/GSE). As there are quite a few 
chiral blocks in table~\ref{tb:toymodelp4} we did not check all of them but concentrated on three cases:
(i) the $\VV_3\oplus \VV_7$ sector for $N_c=10$, 
(ii) the $\VV_3\oplus \VV_7$ sector for $N_c=11$, and 
(iii) the $\VV_0\oplus \VV_4$ sector for $N_c=12$. To identify 
the symmetry classes we again used the probability distribution of the 
ratio of two consecutive level spacings. 
Our numerical results are displayed in figure~\ref{fg:logrp4}, where 
excellent agreement with the respective symmetry classes predicted 
by table~\ref{tb:toymodelp4} is found. 
This corroborates our symmetry classification scheme. 

\begin{figure}[h!]
	\centering
	\includegraphics{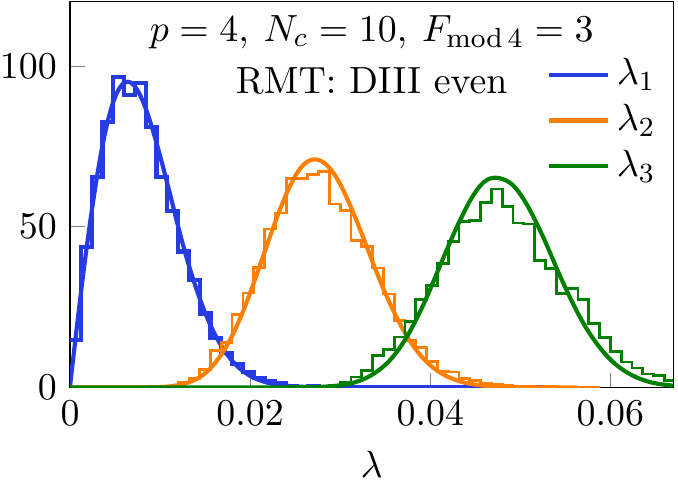}\qquad 
	\includegraphics{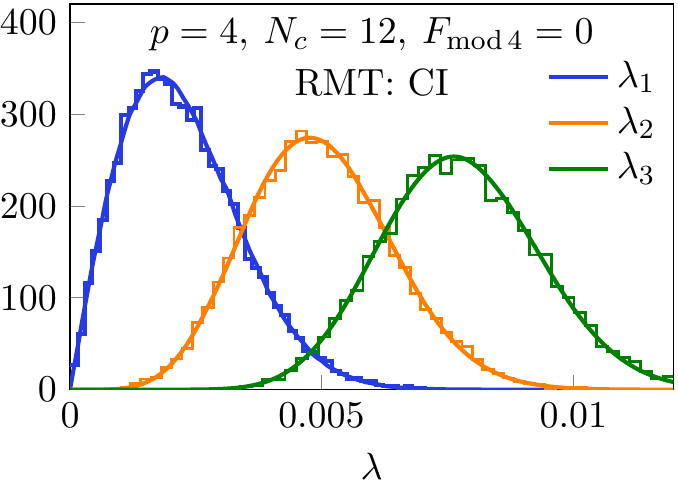}
	\vspace{-7pt}
	\caption{\label{fg:p4hardedgeN10}Distributions of the smallest three 
	eigenvalues of $M$ in \eqref{eq:Mdefin} with $p=4$ and $J=1$ for  
	two sets of $N_c$ and $F$ (mod~4).
	Comparison is made with the predictions (solid lines) of the RMT classes in 
	table~\ref{tb:toymodelp4}. The number of independent 
	random samples is $10^4$ for each plot. The small deviations from RMT are again effects of finite $N_c$.} 
\end{figure}

\paragraph{Universality at the hard edge} \ \\
To check the universality of the level distributions near the origin, 
we have numerically generated $M$ randomly and computed the smallest 3 eigenvalues.  
(In the sector of $F\mequiv 3$ (mod~4) for $N_c=10$, each twofold degenerate pair 
of levels was counted only once.) 
The results shown in figure~\ref{fg:p4hardedgeN10} 
display excellent agreement with RMT as predicted by table~\ref{tb:toymodelp4}.  

\section{\boldmath $\NN=2$ SYK model \label{sc:N2SYK}}
\subsection{Preliminaries}

The $\NN=2$ SYK model \cite{Fu:2016vas,Yoon:2017gut,Peng:2017spg} 
has significantly different properties from its $\NN=1$ cousin. The Hamiltonian is 
defined by $H=\{Q,\bar{Q}\}$ with two supercharges   
\ba
	Q = i \!\!\!\! \sum_{1\leq i<j<k\leq N_c} 
	\!\!\!\! X_{ijk}c_i c_j c_k 
	\qquad \text{and} \qquad 
	\bar{Q} = i \!\!\!\! \sum_{1\leq i<j<k\leq N_c} \!\!\!\!
	\bar{X_{ijk}}\;\bar{c}_i \bar{c}_j \bar{c}_k 
	\label{eq:qqbardef}
\ea
that are nilpotent, $Q^2=\bar{Q}^2=0$, 
where the couplings $X_{ijk}$ are independent complex Gaussian random 
variables obeying $\langle X_{ijk}\bar{X_{ijk}}\rangle=2J/N_c^2$\,. 
Apart from the random disorder, this model is somewhat similar to 
lattice models with exact SUSY \cite{Nicolai:1976xp,Nicolai:1977qx,
Fendley:2002sg,Fendley:2003je,Fendley:2005ae,Eerten2005}. 
The model can be generalized so that $Q$ and $\bar{Q}$ involve 
$\qq$ fermions with $\qq$ odd \cite{Fu:2016vas}. We postpone 
this generic case to section~\ref{sc:genN2q3} 
and for the moment focus on $\qq=3$, i.e., \eqref{eq:qqbardef}.
As for the operator $P$ in \eqref{eq:defP}, we have
\ba
	\label{eq:prm4syk2}
	\begin{tabular}{c|c|cc|l}
		$N_c$ (mod~4) \!\!\! & $P^2$ & & 
		\\\hline 
		0 & $+1$ & $PQ=\bar{Q}P$, & $P\bar{Q}=QP$ & 
		\multirow{4}{*}[0pt]{$\begin{array}{c}\![P,H]=0 \\ 
		\;\text{for~all~}N_c\,. \end{array}$}
		\\
		1 & $+1$ & $PQ=-\bar{Q}P$, & $P\bar{Q}=-QP$ & 
		\\
		2 & $-1$ & $PQ=\bar{Q}P$, & $P\bar{Q}=QP$ &  
		\\
		3 & $-1$ & $PQ=-\bar{Q}P$, & $P\bar{Q}=-QP$ & 
	\end{tabular}
\ea
As shown in \cite{Fu:2016vas,Stanford:2017thb,Mertens:2017mtv}, 
$H$ possesses a number of exactly zero eigenvalues, so SUSY is 
\emph{not} spontaneously broken in contrast to the $\NN=1$ model. 
Moreover, the $\NN=2$ model has $\U(1)$ R-symmetry. $[H,F]=0$ ensures  
that $H$ and $F$ can be diagonalized simultaneously. The total 
Hilbert space $V$ has the structure
\ba
	V=\bigoplus_{f=0}^{N_c}V_f
	\quad\text{with}\quad \dim(V_f)=\binom{N_c}{f}\,,
\ea
where $V_f$ is the eigenspace of $F$ with eigenvalue $f$.
The level density of $H$ in the low-energy limit has been derived analytically 
from the large-$N_c$ Schwarzian theory \cite{Stanford:2017thb,Mertens:2017mtv}, 
whereas analysis of the level statistics and symmetry classification of $H$ based on RMT 
has not yet been done for the $\NN=2$ SYK model. In the remainder of 
this section we fill this gap.

\subsection{\label{sc:naps}Na\"ive approach with partial success}

In this subsection we briefly review a simple approach to the $\NN=2$ model 
that is a natural extrapolation of our treatment for the $\NN=0$ and $1$ 
SYK models but is beset with fatal problems and eventually fails. 
This subsection is included for 
pedagogical reasons and can be skipped by a reader interested only in final results. 

In section~\ref{sc:compsyk} we have reviewed the symmetry properties of 
the $\NN=0$ SYK model with complex fermions, which had the virtue of the 
exactly conserved fermion number, just like the $\NN=2$ SYK model. 
If one were to boldly extrapolate the statements in section~\ref{sc:compsyk} 
to the $\NN=2$ case, one would conclude that the levels of $H$ 
in all $V_f$ except for $V_{N_c/2}$ belong to GUE while those in 
$V_{N_c/2}$ belong to GOE or GSE depending on $P^2=\pm 1$. 
However, numerical analysis of the level correlations clearly reveals disagreement 
with the expected statistics. This failure can be traced back to the fact that 
in this approach all the fine structure of $H$ imposed by $\NN=2$ 
SUSY is neglected. 

So let us change the strategy and try to move along the path 
we have followed in sections~\ref{sc:N1SYK} and \ref{sc:interlude}. 
First of all, note that in the $\NN=2$ SYK model one can write $H=M^2$ 
with a Hermitian operator $M\equiv Q+\bar{Q}$\,. 
Since $M$ preserves $F$ (mod~3) and anticommutes with $(-1)^F$, 
it is useful to divide $V$ into subspaces $\VV_f$ on which $F=f$ (mod~6), i.e.,
\ba
	V=\bigoplus_{f=0}^5\VV_f \quad 
	\text{with}\quad \DD_f \equiv \dim(\VV_f) = 
	\sum_{k=0}^{\lfloor (N_c-f)/6 \rfloor}
	\binom{N_c}{6k+f}\,.
	\label{eq:V6dec}
\ea 
Closed analytic expressions for $\DD_f$ are given in appendix~\ref{ap:Dn}. 
Then $M$ assumes a block-diagonal chiral form $M=\scalebox{0.8}{$\displaystyle 
\bep\0& A_0 \\ A_0^\dagger & \0\eep\oplus 
\bep\0& A_1 \\ A_1^\dagger & \0\eep\oplus 
\bep\0& A_2 \\ A_2^\dagger & \0\eep$}$, where the terms correspond to 
$\VV_0\oplus\VV_3$, $\VV_1\oplus\VV_4$, and $\VV_2\oplus\VV_5$, respectively. 
The spectrum of $M$ has a mirror symmetry for every single realization of $\{X_{ijk}\}$. 
As a consequence, every nonzero eigenvalue of $H$ is at least twofold degenerate. 
From the above structure, a lower bound on the number $N^z$ of exact zero modes of 
$M$ and hence of $H$ can readily be obtained (cf.~appendix~\ref{ap:Dn}) as
\ba
	\label{eq:Nzana}
	N^z \geq \sum_{f=0,1,2}|\DD_{f} - \DD_{f+3}| 
	= 
        \begin{cases}
          4\cdot 3^{N_c/2-1} & \text{for } N_c~\text{even}\,,\\
          2\cdot 3^{(N_c-1)/2} & \text{for } N_c~\text{odd}\,.
        \end{cases}
\ea 
The same bound was obtained via the Witten index in \cite{Sannomiya:2016mnj}.%
\footnote{We emphasize that the extensive number of zero-energy states in this model owes their 
existence to the mismatch of $\DD_f$ and $\DD_{f+3}$ ($f=0,1,2$). 
If one adds an arbitrarily small perturbation that breaks the $\U(1)$ R-symmetry down to $\ZZ_2$, 
the Hamiltonian would lose its triple chiral-block structure and is left with just the two 
eigenspaces of $(-1)^F$, which have equal dimension. Then nothing protects zero modes from 
being lifted and SUSY gets broken, as reported in \cite{Sannomiya:2016wlz,Sannomiya:2016mnj}.}  
In numerical simulations we found that this bound is saturated for $N_c\in\{0,2,3\}$ (mod~4), 
while a strict inequality holds for $N_c=1$ (mod~4) due to the presence of 
$\Or(1)$ ``exceptional'' zero modes \cite{Fu:2016vas,Sannomiya:2016mnj} 
(see also appendix~\ref{ap:NNN}). We will 
explain their origin later. We note in passing that the present argument based on $M$ does not 
tell us how many zero modes exist in each $V_f$.

\begin{figure}[htb]
	\centering
	\includegraphics[height=44mm]{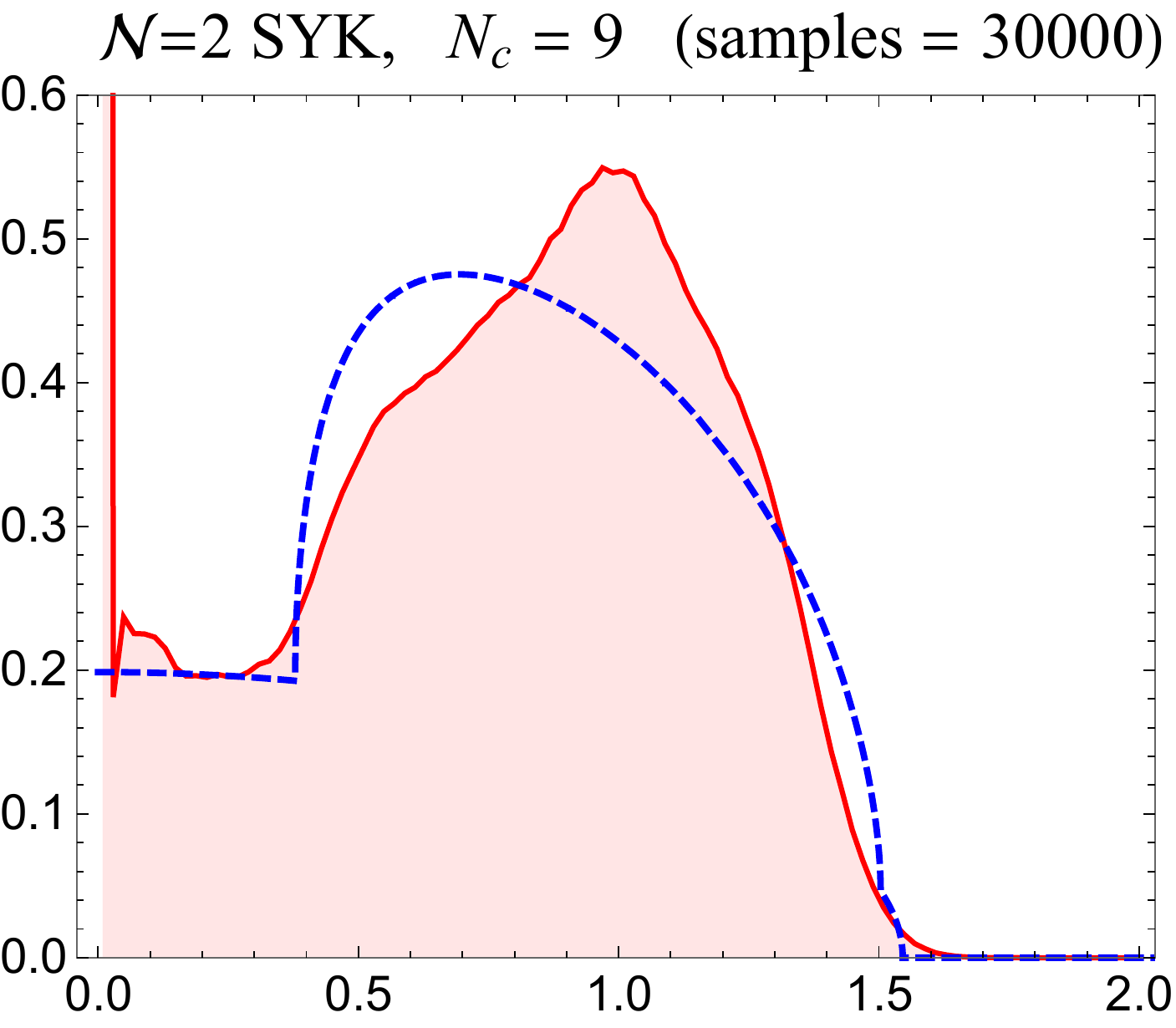}
	\hspace*{15mm}
	\includegraphics[height=44mm]{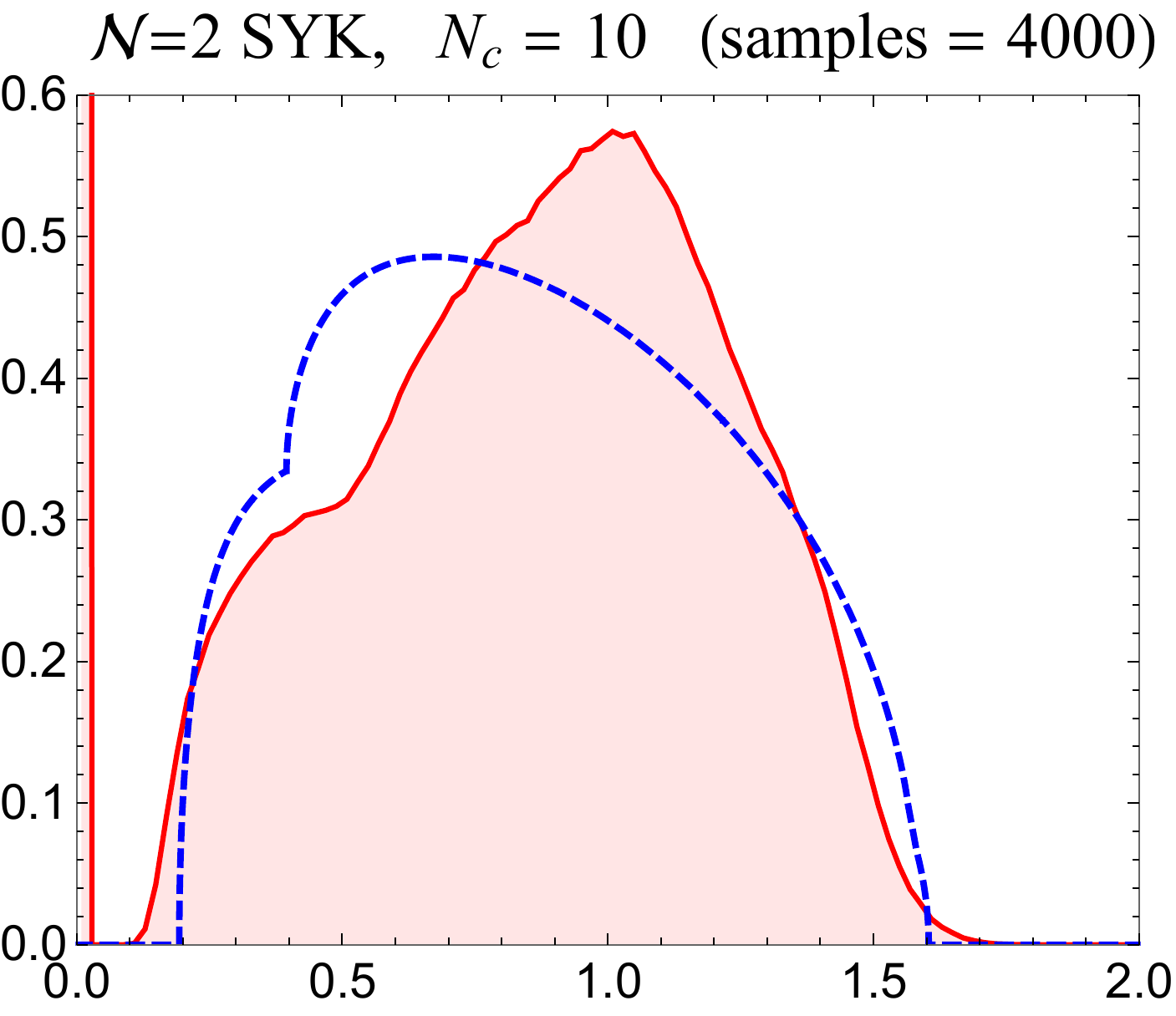}
	\caption{\label{fg:N2SYKglobal}Spectral density of $M=Q+\bar{Q}$ 
	in the $\NN=2$ SYK model from exact diagonalization for $N_c=9$ and 10, averaged 
	over random samples.  
	Since the spectra are symmetric about $0$, only the nonnegative part is shown. 
	The delta peaks at the origin represent exact zero modes, as in figure~\ref{fg:totaldensity910}.
	In both plots $J=1$ and the total density is normalized to 1. 
	The blue dashed lines are the best fits of 
	analytic approximations based on the Mar\v{c}enko-Pastur law.}
\end{figure}

\paragraph{Global spectral density}  \ \\
Utilizing the decomposition of $M$ into three chiral blocks, 
we can derive an approximate analytic formula for the global level density 
based on the Mar\v{c}enko-Pastur law \eqref{eq:Pmpl}, 
repeating the steps that led to \eqref{eq:Pp4appro}. 
(We note that the level densities of $M$ and $H$ are linked by formula 
\eqref{eq:HQrelate}, where $Q$ should be replaced by $M$ here.)
Figure~\ref{fg:N2SYKglobal} displays the numerically obtained 
global spectral density of $M$ for $N_c=9$ and $10$ together with the
analytic approximations obtained by tuning the parameter $\sigma$ for optimal fits. 
The quality of the agreement is worse than for the previous model 
(figure~\ref{fg:totaldensity910}). In particular, the pronounced sharp peak of the density cannot be 
reproduced with the Mar\v{c}enko-Pastur law. This could be an indication that  
the $\NN=2$ SYK model indeed has a more complex structure than the model in 
section~\ref{sc:interlude}. 

In figure~\ref{fg:N2SYKglobal} there is a spectral gap for $N_c=9$ but not for $N_c=10$. 
The peculiar dependence of the level density of $H$ on the parity of $N_c$ was also noted 
in \cite{Stanford:2017thb}. Intriguingly, this can easily be accounted for by the 
fact that a chiral block with $\DD_f=\DD_{f+3}$ is present only for odd $N_c$ 
(cf.~appendix~\ref{ap:Dn}). 
This can be shown by elementary combinatorics.

\paragraph{\boldmath Symmetry of $M$}  \ \\
To classify $M$ based on RMT we can again make use of $P$ and $R \equiv P(-1)^F$ 
in the same way as for the $\NN=1$ SYK model (section~\ref{sc:N1SYK}).  
For $N_c=1$ (mod~4), it can easily be shown that 
$P$ and $R$ map $\VV_f \oplus \VV_{f+3}$ to itself, with 
$f=\scalebox{0.8}{$\displaystyle\left\{\begin{array}{c}2\\1\\0\end{array}\right\}$}$ for 
$N_c=\scalebox{0.8}{$\displaystyle\left\{\begin{array}{c}1\\5\\9\end{array}\right\}$}$ 
\vspace{-5pt}(mod~12). Using \eqref{eq:prm4syk2} one can show\vspace{-3pt}
\begin{equation}
	P^2=+1\,, \quad R^2=-1\,, \quad [R,M]=0\,, \quad \text{and} \quad \{P,M\}=0\,,
\end{equation}  
so $M$ on the corresponding space $\VV_f \oplus \VV_{f+3}$ is classified as
class BdG\,(DIII) with $\beta=4$, according to table~\ref{tb:RMTlist}. Therefore 
every eigenvalue of $M$ must be twofold degenerate. On the other hand, with elementary combinatorics, 
one can show that $\DD_f=\DD_{f+3}=(2^{N_c-1}-1)/3\equiv d_\text{odd}$ 
(cf.~appendix~\ref{ap:Dn}) 
for the three sets of $f$ and $N_c$ specified above. The point is that $d_\text{odd}$ is an odd integer.  
This means that the spectrum of $M$ on $\VV_f \oplus \VV_{f+3}$ 
cannot consist of $d_\text{odd}$ positive levels and $d_\text{odd}$ negative 
levels, since this would contradict the Kramers degeneracy. We conclude that 
$M$ (and $H$) must have at least 2 zero modes in $\VV_f \oplus \VV_{f+3}$. 
This explains why we encounter ``exceptional'' zero modes for $N_c=1$ (mod~4), and
is corroborated by our exact diagonalization analysis of $H$ 
(see appendix~\ref{ap:NNN}).%
\footnote{For $N_c=5,13,17$ we found $2$ exceptional zero modes, while 
only for $N_c=9$ we found $6$ exceptional zero modes, 
in agreement with previous numerical data \cite{Fu:2016vas,Sannomiya:2016mnj}. 
Currently the origin of the 4 additional zero modes is unclear.}  

It turns out, however, that the current approach is incapable of describing 
the actual level structure of $M$ in full detail. For instance, although $M$ in the sector 
$\VV_0\oplus\VV_3$ with $N_c=12$ is classified as class chGOE\,(BDI)$_{\beta=1}$, 
exact diagonalization shows that all nonzero eigenvalues of $M$ 
in this sector are in fact \emph{twofold degenerate}. The reason that the
symmetry classification based on $M$ is doomed to be incomplete is that $M$  
does not manifestly reflect the fermion-number conservation of $H$.  
We have no access to the level statistics in the individual eigenspaces $V_f$ of $F$ 
as long as we see $H$ through the lens of $M$. The upshot is that since 
the structure of the $\NN=2$ SYK model is qualitatively 
different from its cousins with $\NN=0$ and 1 SUSY, we need
an entirely new approach to carry out its symmetry classification. 
This is the subject of the next subsection.

\subsection[Complete classification based on $Q\bar{Q}$ and $\bar{Q}Q$]
{\boldmath Complete classification based on $Q\bar{Q}$ and $\bar{Q}Q$}

Using the nilpotency $Q^2=\bar{Q}^2=0$ one can show 
that $H$, $Q\bar{Q}$, $\bar{Q}Q$ and $F$ all commute with one another, 
so they can be diagonalized simultaneously. Let $\psi$ be an eigenstate  
with $Q\bar{Q}\psi=\E_+\psi$ and $\bar{Q}Q\psi=\E_-\psi$ with $\E_+$, $\E_-\geq 0$. 
Let us assume $\E_+>0$ and $\E_->0$. Then  
\ba
	\begin{split}
		\psi^\dagger \psi & = \frac{1}{\E_+\E_-}(\E_+ \psi)^\dagger \E_- \psi 
		= \frac{1}{\E_+\E_-}(Q\bar{Q}\psi)^\dagger \bar{Q}Q \psi
		\\
		& = \frac{1}{\E_+\E_-} \psi^\dagger Q\bar{Q}^2Q\psi =0\,, 
		\qquad (\because\,\bar{Q}^2=0)
	\end{split}
\ea
implying $\psi$ is a null vector. To resolve this contradiction, 
$\E_+=0$ or $\E_-=0$ must hold for every eigenstate. 
Note that $\E_+=0$ ($\E_-=0$) is equivalent to $\bar{Q}\psi=0$ ($Q\psi=0$) since, e.g., 
$Q\bar{Q}\psi=0$ implies $\psi^\dagger Q\bar{Q}\psi=||\bar{Q}\psi||^2=0$. 
If $\E_+=\E_-=0$, then $\psi$ is a zero mode (ground state) of $H$. Thus 
each subspace $V_f$ of $V$ for given $N_c$ admits 
an orthogonal decomposition 
\ba
	V_f=V_f^{+}\oplus V_f^{-}\oplus V_f^{z}\,, 
\ea
where 
\begin{align}
  V_f^{+}&=\text{Hilbert space spanned by eigenstates $\psi$ with $\bar{Q}\psi\ne0$ and $Q\psi=0$}\,,\notag\\
  V_f^{-}&=\text{Hilbert space spanned by eigenstates $\psi$ with $\bar{Q}\psi=0$ and $Q\psi\ne0$}\,,\\
  V_f^{z}&=\text{Hilbert space spanned by zero modes ($Q\psi=\bar{Q}\psi=0$)}\,.\notag
\end{align}
(In figure~\ref{fg:boxes_Nc15} below we will show a graphical representation of the interrelations of the $V_f^{\pm,z}$.)  
Next we introduce notation for the dimensions of the subspaces,
\ba
	\label{eq:defnnn}
	\begin{split}
           N_f^+ & \equiv \dim(V_f^+)\,, \quad N_f^- \equiv \dim(V_f^-)\,, \quad 
           N_f^z \equiv \dim(V_f^z)\,, 
           \\
           N_f & \equiv\dim(V_f) = N_f^+ + N_f^- + N_f^z =
           \binom{N_c}{f}\,, \quad N^z = \sum_{f=0}^{N_c}N_f^z \,.
	\end{split}
\ea
We choose to keep the $N_c$-dependence of $N_f^{\pm,z}$ implicit to avoid cluttering the notation.  
Using the properties \eqref{eq:prm4syk2} related to $P$ one can verify   
\ba
	\label{eq:NNNre}
	N_f^+ = N_{N_c-f}^-\,, \quad N_f^- = N_{N_c-f}^+\,, \quad 
	N_f^z = N_{N_c-f}^z\,.
\ea
There is yet another important formula for $N_f^\pm$\,. To derive it, we note 
that there is a one-to-one mapping between the bases of $V_f^+$ and those of $V_{f+3}^-$. 
Namely, if $\psi\in V_f^+$ with $Q\bar{Q}\psi=\E\psi$ for $\E>0$, then 
$\psi'\equiv \frac{1}{\sqrt{\E}}\bar{Q}\psi \in V_{f+3}^-$ with $\bar{Q}Q\psi'=\E\psi'$. 
This can be inverted to give $\psi=\frac{1}{\sqrt{\E}}Q\psi'$. Hence 
\ba
	\label{eq:QNNr}
	\left.
	\begin{array}{rl}
		\bar{Q}(V_f^+) & = V_{f+3}^{-} 
		\\
		Q (V_{f+3}^-) & = V_{f}^{+}
	\end{array}
	\right\}
	~\text{for}\;\;0\leq f\leq N_c-3 \quad \text{and}
	\quad N_{f}^{+}=N_{f+3}^{-} \,.
\ea
For convenience we provide tables of the numerical values of $N_f^{\pm,z}$ 
for $3\leq N_c\leq 17$ in appendix~\ref{ap:NNN}. 
They confirm the relations \eqref{eq:NNNre} and \eqref{eq:QNNr}. Explicit analytical formulas for $N_f^{\pm,z}$ will be derived in section~\ref{sec:Nanalytic}.

This concludes the necessary preparations for the ensuing analysis. 
Our strategy in what follows is determined by the observation that $H$ is the sum of two operators that commute with each other.
Therefore we need to classify the symmetries of $H$ on $V_f^+$ and $V_f^-$ 
separately. It is essential to distinguish these eigenspaces 
because they are not mixed by $H$ and the eigenvalues of $H$ on them are, a priori, statistically uncorrelated. 
Na\"ively collecting all eigenvalues of $H$ on $V_f$ leads to incorrect statistics and 
must be avoided.

For generic $f$ and $N_c$\,, there is no antiunitary symmetry that acts within $V_f^\pm$.
$P$ just exchanges $V_f^{+}$ and $V_{N_c-f}^{-}$ (as well as 
$V_f^{-}$ and $V_{N_c-f}^{+}$), which does not impose constraints 
on the level statistics in any of the $V_f^\pm$. 
Therefore the symmetry class of $H$ on $V_f^{\pm}$ is generally GUE.  

\begin{figure}[tbh]
  \centering
  \includegraphics{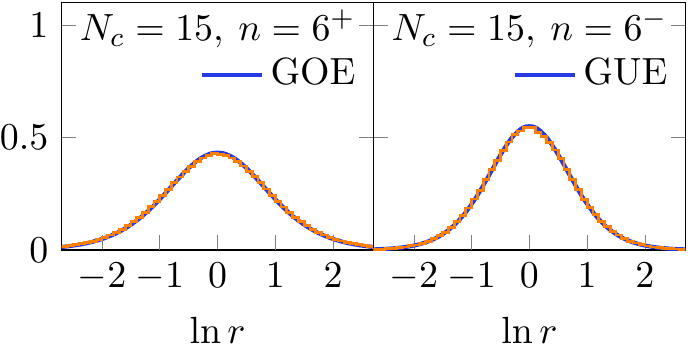}\hfill
  \includegraphics{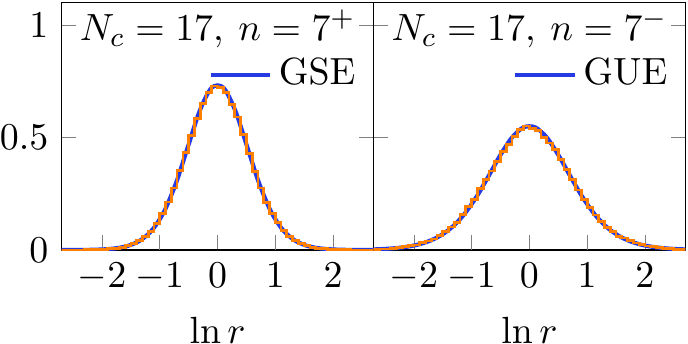}\\[10pt]
  \includegraphics{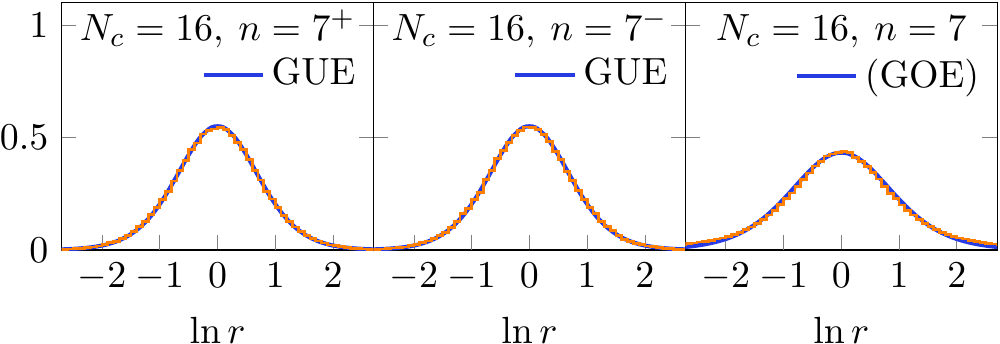}
	\vspace{-4pt}
	\caption{\label{fg:PlnrN2}Distribution of the ratio $r$ of 
        two consecutive level spacings for the $\NN=2$ SYK model with $\qq=3$. 
	The label $n=f^\pm$ $(f=6,7)$ refers to levels of $H$ on $V_f^{\pm}$. 
	The number of realizations used for averaging is $10^3$ for $N_c=15$ and  
	$10^2$ for $N_c=16$ and $17$. The blue lines are surmises for the RMT classes 
	of \eqref{eq:RMTclassN2q3}.  The twofold degeneracy for the GSE case
	was resolved before the statistical analysis. In the right-most plot of $N_c=16$ we show 
	the result obtained by an incorrect analysis, when levels from $V_f^+$ and $V_f^-$ are mixed 
	into a single sequence. 
	Although the result is surprisingly well fitted by the GOE, this is 
	misleading: there is no antiunitary symmetry in this sector. 
	This highlights the danger of inferring the symmetry class from spectra on the full $V_f$.}
\end{figure}

However, when the difference of $f$ and $N_c-f$ is $3$, 
there exists an antiunitary operator that commutes with $H$ and 
maps $V_f^\pm$ to itself. To see this, assume $f+3=N_c-f$ and let $\psi$ be a basis element of $V_f^+$ 
(so that $Q\bar{Q}\psi=\E\psi$ for some $\E>0$). 
Then $\bar{Q}\psi\in V_{f+3}^{-}$, cf.~\eqref{eq:QNNr},
and $P\bar{Q}\psi\in V^+_{f}$, so 
$P\bar{Q}$ is an antilinear operator that acts within $V_f^+$.  
By the same token one can show that $PQ$ maps $V_{f+3}^{-}$ to itself. 
The presence of these operators indicates that the spectra of $H$ on $V_f^+$ 
and $V_{f+3}^-$ in the case $f+3=N_c-f$ belong to either GOE or GSE. If we define 
the canonically normalized operators $P\bar{Q}/\sqrt{H}$ on $V^+_f$ and 
$PQ/\sqrt{H}$ on $V^-_{f+3}$\,, one can show with the help of 
\eqref{eq:prm4syk2} that they are antiunitary and that 
their squares are $\pm\1$, depending on $N_c$ (mod~4).  
This sign determines the symmetry class (GOE/GSE). 
Our conclusions for the $\NN=2$ SYK model with $\qq=3$ are summarized in the following table.  
\ba
	\label{eq:RMTclassN2q3}
	\renewcommand{\arraystretch}{1.3}
	\begin{tabular}{|c||c|c|c|}
		\hline 
		& \cellcolor{cgray} $N_c=0,2$ (mod~4) & 
		\cellcolor{cgray} $N_c=1$ (mod~4) & 
		\cellcolor{cgray} $N_c=3$ (mod~4)
		\\\hhline{|=#=|=|=|}
		$V_f^+$ & {\bf GUE}~ for $\forall f$ & 
		$\begin{array}{cl}
			\text{\textcolor{blue}{\bf GSE}} & ~\text{for $f=\frac{N_c-3}{2}$} \\ 
			\text{{\bf GUE}} & ~\text{for $f\ne\frac{N_c-3}{2}$}
		\end{array}$ 
		& 
		$\begin{array}{cl}
			\text{\alert{\bf GOE}} & ~\text{for $f=\frac{N_c-3}{2}$} \\ 
			\text{{\bf GUE}} & ~\text{for $f\ne\frac{N_c-3}{2}$}
		\end{array}$
		\\\hline 
		$V_f^-$ & {\bf GUE} ~for $\forall f$ & 
		$\begin{array}{cl}
			\text{\textcolor{blue}{\bf GSE}} & ~\text{for $f=\frac{N_c+3}{2}$} \\
			\text{{\bf GUE}} & ~\text{for $f\ne \frac{N_c+3}{2}$}
		\end{array}$ 
		& 
		$\begin{array}{cl}
			\text{\alert{\bf GOE}} & ~\text{for $f=\frac{N_c+3}{2}$} \\
			\text{{\bf GUE}} & ~\text{for $f\ne \frac{N_c+3}{2}$}
		\end{array}$
		\\\hline 
	\end{tabular}
	\renewcommand{\arraystretch}{1}
\ea
This is the main result of this section. 
We have verified our classification by extensive numerical analysis of the spectra of 
$H$ projected to each $V_f$.  
The numerical results shown in figure~\ref{fg:PlnrN2} demonstrate excellent agreement 
with the RMT statistics specified in \eqref{eq:RMTclassN2q3}. Thus, 
as far as one can judge from the short-range correlations of energy levels,  
the $\NN=2$ SYK model exhibits quantum chaos in each eigenspace of $F$ 
to the same extent as its $\NN=0$ and 1 cousins. 

The argument above also clarifies the degeneracy of individual levels of $H$ when 
diagonalized on the whole Hilbert space $V$. In summary, we have found the following:

For $N_c=0,1,2$ (mod~4),
every positive eigenvalue of $H$ is 4-fold degenerate. 
A quadruplet is formed by the set of eigenstates
\ba
  \psi \in V_f^+, \quad \bar{Q}\psi\in V_{f+3}^{-}, \quad P\psi \in V_{N_c-f}^{-}\,,
  \quad \text{and} \quad P\bar{Q}\psi \in V_{N_c-f-3}^{+}
  \label{eq:4sts}
\ea
for $0\leq f\leq N_c-3$\,. The number of quadruplets is $(2^{N_c}-N^z)/4$. 
In particular, for even $N_c$\,, every positive eigenvalue of $H$ on $V_{N_c/2}$ 
is twofold degenerate, because both $\psi$ and $P\psi\in V_{N_c/2}$\,.\footnote{The reader should be cautioned that 
  this degeneracy does \emph{not} mean that $H$ on $V_{N_c/2}$ obeys GSE statistics. 
  Actually, we have two identical copies of the GUE.}

For $N_c=3$ (mod~4), there are
$N_{\scriptscriptstyle(N_c-3)/2}^{+}\, (=N_{\scriptscriptstyle(N_c+3)/2}^{-})$ doublets residing
in the GOE sectors and $(2^{N_c}-N^z-2N_{\scriptscriptstyle(N_c-3)/2}^{+})/4$
quadruplets. The latter consist of the set \eqref{eq:4sts}
subject to the condition that $f\ne (N_c-3)/2$.
	
\subsection[Analytical formulas for $N_f^\pm$ and $N_f^z$]
{\boldmath Analytical formulas for $N_f^\pm$ and $N_f^z$}
\label{sec:Nanalytic}

\begin{figure}[b]
	\centering 
	\includegraphics[width=.88\textwidth]{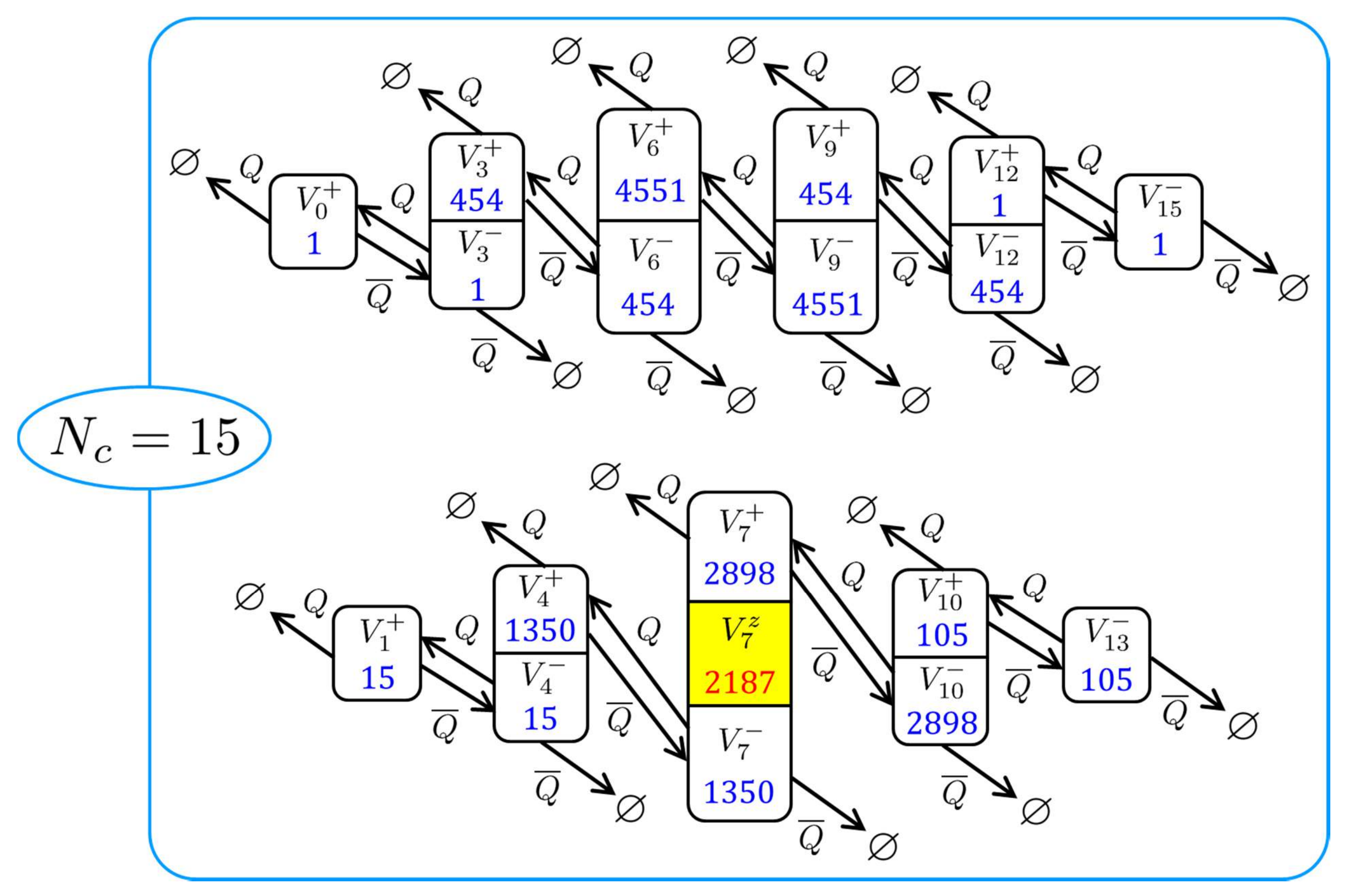}\qquad 
	\vspace{-6pt}
	\caption{\label{fg:boxes_Nc15}Relations among the Hilbert spaces 
	with $F=0$ and $1$ (mod~3) for $N_c=15$. The numbers shown are the
	dimensions of the corresponding subspaces of $V$. 
	Arrows to the symbol $\varnothing$ (empty set) are shown to 
	emphasize the nilpotency $Q^2={\bar{Q}}^2=0$.}
\end{figure}

Up to now we have not mentioned how to compute $N_f^{\pm}$ explicitly for given $f$ and $N_c$\,. 
Actually this proves to be a straightforward (albeit tedious) task if we posit the following premise:
\vspace{-3pt}
\begin{equation}
	\label{eq:zeroAnsatz}
	\begin{minipage}{.83\textwidth}
		\emph{For any $N_c\geq 3$, all exact zero modes of $H$ reside in $V_f$ with $|f-N_c/2|\leq 3/2$,
		where the equality holds only for exceptional zero modes that occur 
		when} $N_c=1$ (mod~4).\footnotemark
	\end{minipage}
	\vspace{-3pt}
\end{equation}
\footnotetext{The origin of these exceptional zero modes was explained in 
section~\ref{sc:naps}.}%
This rather strong condition on the ground states of $H$ is not only corroborated by detailed numerical 
simulations (see appendix~\ref{ap:NNN} and \cite{Fu:2016vas}) but also 
derived from the Schwarzian effective theory valid in the large-$N_c$ and low-energy 
limit \cite{Stanford:2017thb,Mertens:2017mtv}. If \eqref{eq:zeroAnsatz} is 
accepted, one can fully clarify the relation of Hilbert spaces linked by $\bar{Q}$ 
as in table~\ref{tb:Vseq}. The sequences tabulated there are \emph{exact sequences} in the terminology of mathematics, 
in the sense that the kernel of $\bar{Q}$ acting on $V_f$ coincides exactly with the image of $V_{f-3}$ by $\bar{Q}$. 
Two examples of these sequences, extended up to $V_{N_c}$, 
are graphically illustrated in figure~\ref{fg:boxes_Nc15} for $N_c=15$. 

\begin{table}[t]
	\caption{\label{tb:Vseq}Exact sequences of the Hilbert spaces generated by the linear map $\bar{Q}$.   
	Complementary exact sequences descending from $V_{N_c}$, 
	$V_{N_c-1}$, and $V_{N_c-2}$ by way of $Q$ can be obtained 
	by applying the particle-hole operator $P$ to 
	the sequences in the table. 
	The spaces $\alert{V}$ contain an exponentially large number of ``typical'' zero modes, see \eqref{eq:Nzana}. 
	The spaces ${\color{blue}V^*}$ contain no zero modes for $N_c=3$ (mod~4), or 
	$1$ or $3$ ``exceptional'' zero modes for $N_c=1$ (mod~4).}
	\vspace{\baselineskip}
	\centering 
	\begin{tabular}{|c|c||c|c|}
		\hline
		\!\!\!\!$\begin{array}{c}N_c\vspace{-5pt}\\\text{\footnotesize (mod~6)}\end{array}$\!\!\!\! 
		& Exact Sequence & 
		\!\!\!\!$\begin{array}{c}N_c\vspace{-5pt}\\\text{\footnotesize (mod~6)}\end{array}$\!\!\!\! 
		& Exact Sequence
		\\\hhline{|=|=#=|=|}
		0 & $\displaystyle\begin{array}{l} 
			V_0 \xrightarrow{\bar{Q}} V_3 \xrightarrow{\bar{Q}} \cdots \xrightarrow{\bar{Q}} \alert{V_{N_c/2}}
			\\ 
			V_1 \xrightarrow{\bar{Q}} V_4 \xrightarrow{\bar{Q}} \cdots \xrightarrow{\bar{Q}} \alert{V_{N_c/2+1}}
			\\ 
			V_2 \xrightarrow{\bar{Q}} V_5 \xrightarrow{\bar{Q}} \cdots \xrightarrow{\bar{Q}} \alert{V_{N_c/2-1}}
		\end{array}$ 
		& 3 & 
		$\displaystyle\begin{array}{l} 
			V_0 \xrightarrow{\bar{Q}} V_3 \xrightarrow{\bar{Q}} \cdots \xrightarrow{\bar{Q}} {\color{blue}V_{(N_c-3)/2}^*}
			\\ 
			V_1 \xrightarrow{\bar{Q}} V_4 \xrightarrow{\bar{Q}} \cdots \xrightarrow{\bar{Q}} \alert{V_{(N_c-1)/2}}
			\\ 
			V_2 \xrightarrow{\bar{Q}} V_5 \xrightarrow{\bar{Q}} \cdots \xrightarrow{\bar{Q}} \alert{V_{(N_c+1)/2}}
			\raisebox{-7pt}{}
		\end{array}$ 
		\\\hline
		1 & 
		$\displaystyle\begin{array}{l} 
			V_0 \xrightarrow{\bar{Q}} V_3 \xrightarrow{\bar{Q}} \cdots \xrightarrow{\bar{Q}} \alert{V_{(N_c-1)/2}}
			\\ 
			V_1 \xrightarrow{\bar{Q}} V_4 \xrightarrow{\bar{Q}} \cdots \xrightarrow{\bar{Q}} \alert{V_{(N_c+1)/2}}
			\\ 
			V_2 \xrightarrow{\bar{Q}} V_5 \xrightarrow{\bar{Q}} \cdots \xrightarrow{\bar{Q}} {\color{blue}V_{(N_c-3)/2}^*}
		\end{array}$
		& 4 & $\displaystyle\begin{array}{l} 
			V_0 \xrightarrow{\bar{Q}} V_3 \xrightarrow{\bar{Q}} \cdots \xrightarrow{\bar{Q}} \alert{V_{N_c/2+1}}
			\\ 
			V_1 \xrightarrow{\bar{Q}} V_4 \xrightarrow{\bar{Q}} \cdots \xrightarrow{\bar{Q}} \alert{V_{N_c/2-1}}
			\\ 
			V_2 \xrightarrow{\bar{Q}} V_5 \xrightarrow{\bar{Q}} \cdots \xrightarrow{\bar{Q}} \alert{V_{N_c/2}}
			\raisebox{-7pt}{}
		\end{array}$
		\\\hline
		2 & $\displaystyle\begin{array}{l} 
			V_0 \xrightarrow{\bar{Q}} V_3 \xrightarrow{\bar{Q}} \cdots \xrightarrow{\bar{Q}} \alert{V_{N_c/2-1}}
			\\ 
			V_1 \xrightarrow{\bar{Q}} V_4 \xrightarrow{\bar{Q}} \cdots \xrightarrow{\bar{Q}} \alert{V_{N_c/2}}
			\\ 
			V_2 \xrightarrow{\bar{Q}} V_5 \xrightarrow{\bar{Q}} \cdots \xrightarrow{\bar{Q}} \alert{V_{N_c/2+1}}
		\end{array}$
		& 5 & 
		$\displaystyle\begin{array}{l} 
			V_0 \xrightarrow{\bar{Q}} V_3 \xrightarrow{\bar{Q}} \cdots \xrightarrow{\bar{Q}} \alert{V_{(N_c+1)/2}}
			\\ 
			V_1 \xrightarrow{\bar{Q}} V_4 \xrightarrow{\bar{Q}} \cdots \xrightarrow{\bar{Q}} {\color{blue}V_{(N_c-3)/2}^*}
			\\ 
			V_2 \xrightarrow{\bar{Q}} V_5 \xrightarrow{\bar{Q}} \cdots \xrightarrow{\bar{Q}} \alert{V_{(N_c-1)/2}} 
			\raisebox{-7pt}{}
		\end{array}$ 
		\\\hline 
	\end{tabular}
\end{table}

Although we do not provide a rigorous proof of \eqref{eq:zeroAnsatz}, 
there is a heuristic argument to convince oneself that \eqref{eq:zeroAnsatz} is correct. 
Let us consider a sequence $\cdots \xrightarrow{\bar{Q}}V_f 
\xrightarrow{\bar{Q}} V_{f+3}\xrightarrow{\bar{Q}}\cdots$ with 
$\dim(V_f)<\dim(V_{f+3})$. If $\bar{Q}$ in the middle were a completely random linear map, 
it is a matrix of size $\dim(V_f)\times \dim(V_{f+3})$ whose rank is almost surely $\dim(V_f)$ 
(in the absence of fine-tuning or a special symmetry).  
This is of course an oversimplification for $\bar{Q}$, because it is not a generic 
linear map but a \emph{nilpotent} map. Taking this into account, let us next view $\bar{Q}$ 
as a random matrix of size $\dim\big[V_f\setminus \bar{Q}(V_{f-3})\big]\times \dim(V_{f+3})$, 
where the trivial kernel $\bar{Q}(V_{f-3})$ has been left out. Then the rank of $\bar{Q}$ is almost surely 
$\dim\big[V_f\setminus \bar{Q}(V_{f-3})\big]$, i.e., there is no ``nontrivial'' zero mode of $\bar{Q}$ 
in $V_f$. This argument may be repeated along the sequence 
as long as the condition $\dim(V_f)<\dim(V_{f+3})$ is fulfilled. 
A completely parallel argument can also be given for a ``descending'' sequence $\cdots \xleftarrow{Q}V_f 
\xleftarrow{Q}V_{f+3}\xleftarrow{Q}\cdots$ with $\dim(V_f)>\dim(V_{f+3})$. By pinching 
the sequence from both ends like this, we find at the end of the day that all zero modes $(Q\psi=\bar{Q}\psi=0)$ 
must be concentrated in the subspace $V_f$ with the \emph{largest dimension} in the sequence. 
This is equivalent to the condition \eqref{eq:zeroAnsatz}. 

Now it is straightforward to work out $N_f^{\pm}$. 
Let us begin with the case of even $N_c$\,. First, for $3\leq f<N_c/2-1$, $V_f$ does not contain 
zero modes under the assumption \eqref{eq:zeroAnsatz}. Hence, with the 
help of \eqref{eq:QNNr}, we find
\ba
	N_f^+ =  \binom{N_c}{f}
	- N_f^- = \binom{N_c}{f} - N_{f-3}^{+}\,.
\ea
This recursion relation for $N_f^+$ is to be solved with the initial conditions
$N_0^+=1$, $N_1^+=N_c$, and $N_2^+=N_c(N_c-1)/2$. The result reads
\ba
	N_f^+ & = N_{N_c-f}^- 
	= (-1)^{f+f_0}N_{f_0}^+ + (-1)^f \sum_{n=1}^{(f-f_0)/3} 
	(-1)^{3n+f_0}
	\binom{N_c}{3n+f_0} \,,
	\label{eq:54uwjfk}
	\\
	N_f^- & = N_{N_c-f}^+ = 
	\binom{N_c}{f} - N_f^+ 
	\,,
	\label{eq:nrq489fq}
\ea
where $f_0\equiv f-3\lfloor f/3 \rfloor\in\{0,1,2\}$. 
Equation~\eqref{eq:NNNre} was used in the first equalities of 
\eqref{eq:54uwjfk} and \eqref{eq:nrq489fq}.  
These formulas hold in the range $0\leq f<N_c/2-1$. 
We verified \eqref{eq:54uwjfk} numerically for $N_c$ up to $17$.

Finally, to derive $N_f^{\pm}$ for $f$ close to $N_c/2$\,, we need to know 
$N_f^z$\,. Recalling the premise \eqref{eq:zeroAnsatz} and 
the fact that the inequality \eqref{eq:Nzana} 
is saturated except when $N_c=1$ (mod~4) (see appendix~\ref{ap:Dn} for $\DD_f$ and section~\ref{sc:naps} for the origin of the 
$1$ or $3$ ``exceptional'' zero modes in this case), 
we readily arrive at the following summary:
\setlength{\tabcolsep}{3pt}
\ba
	\label{eq:zerotbl}
	\scalebox{0.9}{\noindent
	\begin{tabular}{|c|c|c|}
		\hline 
		\cellcolor{cgray} $N_c=0,2$ (mod~4) & 
		\cellcolor{cgray} $N_c=1$ (mod~4) & 
		\cellcolor{cgray} $N_c=3$~(mod~4)
		\\\hline
		&&\\[-1em] 
		$N_f^z=\left\{\begin{array}{ll}
			2\cdot 3^{N_c/2-1}\,, & \displaystyle f = \frac{N_c}{2}
			\vspace{4pt}\\
			3^{N_c/2-1}\,, & \displaystyle f = \frac{N_c}{2} \pm 1 \!\!
			\\
			0\,, & \text{otherwise}
		\end{array}\right.$ 
		& 
		$N_f^z=\left\{\begin{array}{ll}
			3^{(N_c-1)/2}\,, & \displaystyle f = \frac{N_c\pm 1}{2}\!\!
			\vspace{4pt}\\
			1~\text{or}~3\,, & \displaystyle f = \frac{N_c\pm 3}{2}\!\!
			\\
			0\,, & \text{otherwise}
		\end{array}\right.$
		&
		$N_f^z=\left\{\begin{array}{ll}
			3^{(N_c-1)/2}\,, & \displaystyle f = \frac{N_c\pm 1}{2}\!\!
			\\
			0\,, & \text{otherwise}
		\end{array}\right.$ 
		\\\hline 
	\end{tabular}
	}
\ea
\setlength{\tabcolsep}{6pt}\noindent 
which fully agrees with numerical results in \cite{Fu:2016vas}. 
This input should be plugged into 
\ba
	N^+_f = \binom{N_c}{f}
	- N^+_{f-3} - N_f^z 
	\quad \text{and}\quad N_f^- = N_{f-3}^+ 
	\qquad \text{for}\quad f=\frac{N_c}{2},\,\frac{N_c}{2}\pm 1\,,
\ea
where $N^+_{f-3}$ has been obtained by \eqref{eq:54uwjfk}. 
This completes our discussion of even $N_c$\,. 

For odd $N_c$, \eqref{eq:54uwjfk} and \eqref{eq:nrq489fq} still hold 
in the range $0\leq f<(N_c-3)/2$ (see table~\ref{tb:Vseq}). For $f$ near 
$N_c/2$ we only have to substitute \eqref{eq:zerotbl} into 
\ba
	N^+_f = \binom{N_c}{f}
	- N^+_{f-3} - N_f^z 
	\quad \text{and}\quad N_f^- = N_{f-3}^+ 
	\qquad \text{for}\quad f=\frac{N_c \pm 1}{2},\,\frac{N_c\pm 3}{2} \,. 
\ea
The numerical results in appendix~\ref{ap:NNN} agree with the formulas derived in this subsection.

\subsection[Generalization to $\qq>3$]
{\boldmath\label{sc:genN2q3}Generalization to $\qq>3$}

\begin{figure}[t]
	\centering 
	\includegraphics{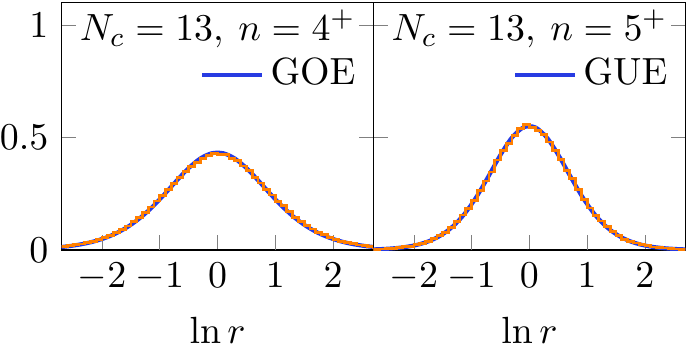}\hspace*{5mm} 
	\includegraphics{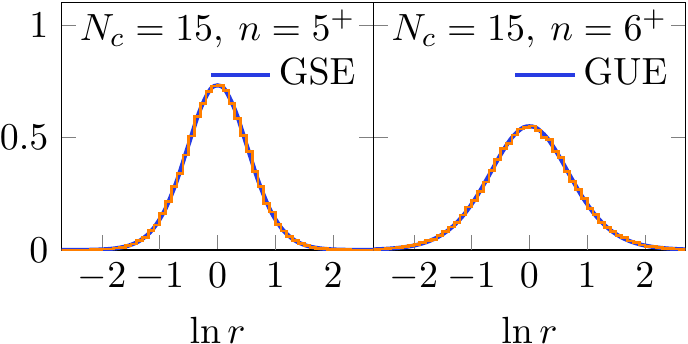}
	\vspace{-5pt}
	\caption{\label{fg:PlnrN2q5}Same as figure~\ref{fg:PlnrN2} 
	but for $\qq=5$ and compared with the surmises of the
	RMT classes in table~\eqref{eq:classN2generalq}. 
	The number of realizations used for averaging is $10^3$ for $N_c=13$ and  
	$10^2$ for $N_c=15$.}
\end{figure}

We now generalize the preceding classification scheme to 
the $\NN=2$ SYK model with $H=\{Q,\bar{Q}\}$ and $\qq$ complex 
fermions in the supercharge, where $\qq$ is odd, i.e.,
\ba
	Q = i^{(\qq-1)/2} \!\!\!\!\!\!\!\!\! \sum_{1\leq i_1<\cdots<i_{\qq}\leq N_c} 
	\!\!\!\!\! X_{i_1 i_2 \cdots i_{\qq}} c_{i_1} c_{i_2} \cdots c_{i_\qq} 
	\quad \text{and} \quad 
	\bar{Q} = i^{(\qq-1)/2} \!\!\!\!\!\!\!\!\! \sum_{1\leq i_1<\cdots<i_{\qq}\leq N_c}
	\!\!\!\!\! \bar{X_{i_1 i_2 \cdots i_{\qq}}} \bar{c}_{i_1} \bar{c}_{i_2} \cdots \bar{c}_{i_\qq} \,.
\ea
This is a counterpart of \eqref{eq:QdefN1} with $\NN=1$. 
For $\qq=3$ it reverts to \eqref{eq:qqbardef}. 
The tables \eqref{eq:prm4syk2} and \eqref{eq:RMTclassN2q3} 
for $\qq=3$ are now generalized to
\ba
	\begin{tabular}{c|c|cc|l}
		$N_c$ (mod~4) \!\!\! & $P^2$ & & 
		\\\hline 
		&&&\\[-1.2em]
		0 & $+1$ & $PQ=(-1)^{\frac{\qq+1}{2}}\bar{Q}P$, & 
		$P\bar{Q}=(-1)^{\frac{\qq+1}{2}}QP$ & 
		\multirow{4}{*}[-5pt]{$\begin{array}{c}\![P,H]=0 \\ 
		\;\text{for~all~}N_c\,. \end{array}$}
		\\
		&&&\\[-1.2em]
		1 & $+1$ & $PQ=(-1)^{\frac{\qq-1}{2}}\bar{Q}P$, & 
		$P\bar{Q}=(-1)^{\frac{\qq-1}{2}}QP$ & 
		\\
		&&&\\[-1.2em]
		2 & $-1$ & $PQ=(-1)^{\frac{\qq+1}{2}}\bar{Q}P$, & 
		$P\bar{Q}=(-1)^{\frac{\qq+1}{2}}QP$ &  
		\\
		&&&\\[-1.2em]
		3 & $-1$ & $PQ=(-1)^{\frac{\qq-1}{2}}\bar{Q}P$, & 
		$P\bar{Q}=(-1)^{\frac{\qq-1}{2}}QP$ & 
	\end{tabular}
\ea
and 
\ba
	\label{eq:classN2generalq}
	\!\!\!\!\!\!\!\scalebox{0.9}{
	\renewcommand{\arraystretch}{1.3}
	\begin{tabular}{|c||c|c|c|}
		\hline 
		& \cellcolor{cgray} $\begin{array}{c}
			\!\!\! N_c=0,\,2 \!\!\! \\[-7pt] (\mathrm{mod}~4)\!
		\end{array}$ & 
		\cellcolor{cgray} $N_c=1$ (mod~4) & 
		\cellcolor{cgray} $N_c=3$ (mod~4)
		\\\hhline{|=#=|=|=|}
		$V_f^+$ & \!\!\!$\begin{array}{c}
			{\bf GUE} 
			\\[-5pt]
			\mathrm{for}~\forall f
		\end{array}$\!\!\! & 
		$\begin{array}{cl}
			\left.
			\begin{array}{c}
				\text{\alert{\bf GOE}}~~
				{\small \begin{matrix}\text{if $\qq=1$}\\[-6pt]\text{(mod~4)}\end{matrix}}
				\\
				\text{\textcolor{blue}{\bf GSE}}~~
				{\small \begin{matrix}\text{if $\qq=3$}\\[-6pt]\text{(mod~4)}\end{matrix}}
			\end{array}
			\right\}
			& \text{for $f=\frac{N_c-\qq}{2}$} \\ 
			\text{{\bf GUE}} & \text{for $f\ne\frac{N_c-\qq}{2}$}
		\end{array}$ 
		& 
		$\begin{array}{cl}
			\left.
			\begin{array}{c}
				\text{\textcolor{blue}{\bf GSE}}~~
				{\small \begin{matrix}\text{if $\qq=1$}\\[-6pt]\text{(mod~4)}\end{matrix}}
				\\
				\text{\alert{\bf GOE}}~~
				{\small \begin{matrix}\text{if $\qq=3$}\\[-6pt]\text{(mod~4)}\end{matrix}}
			\end{array}
			\right\}
			& \text{for $f=\frac{N_c-\qq}{2}$} \\ 
			\text{{\bf GUE}} & \text{for $f\ne\frac{N_c-\qq}{2}$}
		\end{array}$
		\\\hline 
		$V_f^-$ & \!\!\!$\begin{array}{c}
			{\bf GUE} 
			\\[-5pt]
			\mathrm{for}~\forall f
		\end{array}$\!\!\! & 
		$\begin{array}{cl}
			\left.
			\begin{array}{c}
				\text{\alert{\bf GOE}}~~
				{\small \begin{matrix}\text{if $\qq=1$}\\[-6pt]\text{(mod~4)}\end{matrix}}
				\\
				\text{\textcolor{blue}{\bf GSE}}~~
				{\small \begin{matrix}\text{if $\qq=3$}\\[-6pt]\text{(mod~4)}\end{matrix}}
			\end{array}
			\right\}
			& \text{for $f=\frac{N_c+\qq}{2}$} \\
			\text{{\bf GUE}} & \text{for $f\ne \frac{N_c+\qq}{2}$}
		\end{array}$ 
		& 
		$\begin{array}{cl}
			\left.
			\begin{array}{c}
				\text{\textcolor{blue}{\bf GSE}}~~
				{\small \begin{matrix}\text{if $\qq=1$}\\[-6pt]\text{(mod~4)}\end{matrix}}
				\\
				\text{\alert{\bf GOE}}~~
				{\small \begin{matrix}\text{if $\qq=3$}\\[-6pt]\text{(mod~4)}\end{matrix}}
			\end{array}
			\right\} 
			& \text{for $f=\frac{N_c+\qq}{2}$} \\
			\text{{\bf GUE}} & \text{for $f\ne \frac{N_c+\qq}{2}$}
		\end{array}$
		\\\hline 
	\end{tabular}
	\renewcommand{\arraystretch}{1}
	}\hspace{-15pt}
\ea
respectively. We numerically tested this table via exact diagonalization of $H$. 
Figure~\ref{fg:PlnrN2q5} shows superb agreement between the numerical data and RMT. 

We also analyzed the dimensions $N_f^{\pm,z}$ of the subspaces, for which formulas similar
to those in section~\ref{sec:Nanalytic} can be derived.
For $\qq=5$, we have numerically confirmed up to $N_c=17$ that all exact zero modes of $H$ 
reside in $V_f$ with $|f-N_c/2|\leq 5/2$.  
The last inequality is saturated only for $N_c=7$ and $11$ by just 2 zero modes in each case. 
This is not only consistent with our heuristic argument in section~\ref{sec:Nanalytic} 
but also conforms to the claim at large $N_c$ \cite{Stanford:2017thb,Mertens:2017mtv} 
that all zero modes should satisfy $|f-N_c/2|<\qq/2$. In the regime $N_c\gg 1$ 
one can ignore $\Or(1)$ exceptional zero modes and the strict inequality may be justified. 

\section{\label{sc:conclusion}Conclusions}

In this paper we have completed the symmetry classification of 
SYK models with $\NN=0$, 1, and 2 SUSY on the basis of the Altland-Zirnbauer theory 
of random matrices (table~\ref{tb:RMTlist}). The symmetry classes of RMT not only tell us the level degeneracies 
of the Hamiltonian but also offer a diagnostic tool of quantum chaos through 
level correlations in the bulk of the spectrum. Furthermore, when the spectral mirror symmetry is present, RMT precisely 
predicts universal level correlation functions in the vicinity of the origin (also known as \emph{hard edge} or \emph{microscopic domain} \cite{Verbaarschot:2000dy}). 
The present work can be viewed as a generalization of preceding works 
\cite{You:2016ldz,Garcia-Garcia:2016mno,Cotler:2016fpe,Garcia-Garcia:2017pzl,Li:2017hdt} 
that analyzed the level statistics of the $\NN=0$ and $1$ SYK models solely with a 4-body interaction.%
\footnote{A notable exception is \cite{You:2016ldz}, which also considered $4k$-body interactions 
with $k\in\mathbb{N}$.} Our new results include the following:
\begin{enumerate}[leftmargin=7mm]
	\item 
	The symmetry classification of the $\NN=0$ SYK model was 
	given for a Hamiltonian with the most generic $q$-body interaction. The result, summarized in 
	tables~\ref{tb:rmtSYK00} and \ref{tb:rmtSYK01}, includes the RMT classes 
	C and D that did not show up in the preceding classification of 
	\cite{You:2016ldz,Garcia-Garcia:2016mno,Cotler:2016fpe,Garcia-Garcia:2017pzl,Li:2017hdt}.  
	Our results were corroborated by detailed numerics (figure~\ref{fg:N0q6logr}). 
	\item
	We numerically compared the smallest eigenvalue distributions in the $\NN=0$ SYK model 
	with $q=6$ with the RMT predictions of class C and D, finding excellent agreement (figure \ref{fg:PminN0}). 
	\item
	The symmetry classification of the $\NN=1$ SYK model was 
	given for a supercharge with the most generic interaction of $\qq$ Majorana fermions 
	(tables~\ref{tb:RMTQ00} and \ref{tb:RMTQ01}). This extends \cite{Li:2017hdt} 
	which investigated only $\qq=3$. Our results were corroborated by detailed numerics (figure~\ref{fg:Plnr_N1}).
	\item 
	We numerically compared the smallest eigenvalue distributions in the $\NN=1$ SYK model 
	with $\qq=3$ and $5$ with the RMT predictions, finding excellent agreement 
	(figures \ref{fg:PminN1q3} and \ref{fg:PminN1q5}). This confirms the hard-edge universality of 
	the $\NN=1$ SYK model for the first time and is relevant for the thermodynamics 
	of this model at low temperatures comparable to the energy scale of the SUSY breaking. 
	\item 
	We proposed an intriguing new SYK-type model which lacks SUSY but whose Hamiltonian is 
	semi-positive definite and has an extensive number of zero-energy states (section~\ref{sc:interlude}). 
	The symmetry classification based on RMT was provided, and a detailed numerical analysis 
	of the spectra both in the bulk and near the origin was performed, resulting in
	agreement with the RMT predictions. 
	\item 
	We completed the RMT classification of the $\NN=2$ SYK model for the first time.  
	This model is qualitatively different from its $\NN=0$ and $1$ cousins in various aspects.
	It is a model of complex fermions rather than Majorana fermions, and it has a $\U(1)$ R-symmetry. 
	The symmetry classification of this model is nontrivial because the structure of its Hilbert space is far more complex 
	(see figure~\ref{fg:boxes_Nc15} for an example)
	than that of the $\NN=0$ SYK model with complex fermions considered previously 
	in \cite{Sachdev:2015efa,You:2016ldz,Fu:2016yrv,Davison:2016ngz,Bulycheva2017}. 
	Our main results, summarized in table \eqref{eq:RMTclassN2q3}  
	for $\qq=3$ and in table \eqref{eq:classN2generalq} for general odd $\qq$, 
	are strongly supported by intensive numerics, as shown in 
	figure~\ref{fg:PlnrN2} (for $\qq=3$) and figure~\ref{fg:PlnrN2q5} (for $\qq=5$). 
	\item 
	In section~\ref{sc:naps} we succeeded in giving a logical explanation 
	for the curious fact \cite{Fu:2016vas,Sannomiya:2016mnj} that, in the $\NN=2$ 
	SYK model, the number of zero-energy ground states exactly agrees with the 
	lower bound from the Witten index in some
	cases but not in other cases. In short, this is due to the dichotomy between 
	the odd dimensionality of the Hilbert space and Kramers degeneracy.
\end{enumerate}
This work can be extended in several directions. First, our analysis of 
spectral properties of the Hamiltonian could be further deepened by 
using probes that are sensitive to long-range correlations of energy levels, 
like the level number variance $\Sigma^2(L)$ and 
the spectral rigidity $\Delta_3(L)$ \cite{Guhr:1997ve,Mehta_book}. 
Investigating the spectral form factor of the $\NN=2$ SYK model and making a quantitative 
comparison with RMT along the lines of \cite{Cotler:2016fpe} 
is another future direction, although physical interpretation of the ramp, dip, etc., of 
the spectral form factor as a signature of quantum chaos is rather subtle \cite{Balasubramanian:2016ids}. 
Finally, we note that there is no analytical result for the global spectral density 
of the $\NN=1$ and $2$ SYK models, although an accurate formula is 
already known for the $\NN=0$ model 
\cite{Garcia-Garcia:2016mno,Cotler:2016fpe,Garcia-Garcia:2017pzl}. 
We wish to address some of these problems in the future.

\acknowledgments 
TK was supported by the RIKEN iTHES project. 
TW was supported in part by the German Research Foundation (DFG) in the
framework of SFB/TRR-55.

\appendix
\section{\boldmath $\DD_f$ in the $\NN=2$ SYK model \label{ap:Dn}}

In this appendix we display short convenient 
expressions for $\DD_f$ as defined in \eqref{eq:V6dec}  
for the $\NN=2$ SYK model with $\qq=3$. 
For simplicity we denote $N_c$ by $N$ in this appendix. Then 
\ba
	\DD_0 & = \frac{1}{6}\kkakko{
		2^N + 2 \cdot 3^{N/2} \cos {\frac{N\pi}{6}} 
		+ 2 \cos {\frac{N\pi}{3}}
	},
	\\
	\DD_1 & = \frac{1}{6}\kkakko{
		2^N - 2 \cdot 3^{N/2}\cos {\frac{(N+4)\pi}{6}} 
		+ 2 \cos {\frac{(N-2)\pi}{3}}
	},
	\\
	\DD_2 & = \frac{1}{6}\kkakko{
		2^N + 2 \cdot 3^{N/2}\cos {\frac{(N-4)\pi}{6}}
		+ 2 \cos {\frac{(N+2)\pi}{3}}
	},
	\\
	\DD_3 & = \frac{1}{6}\kkakko{
		2^N - 2 \cdot 3^{N/2} \cos {\frac{N\pi}{6}} 
		+ 2 \cos {\frac{N\pi}{3}}
	},
	\\
	\DD_4 & = \frac{1}{6}\kkakko{
		2^N + 2 \cdot 3^{N/2}\cos {\frac{(N+4)\pi}{6}} 
		+ 2 \cos {\frac{(N-2)\pi}{3}}
	},
	\\
	\DD_5 & = \frac{1}{6}\kkakko{
		2^N - 2 \cdot 3^{N/2}\cos {\frac{(N-4)\pi}{6}}
		+ 2 \cos {\frac{(N+2)\pi}{3}}
	}.
\ea

\section{\boldmath\label{ap:NNN}Dimensions of Hilbert spaces for $\NN=2$}

In this appendix we present tables of the $N_f^{\pm,z}$ defined in \eqref{eq:defnnn}
for the $\NN=2$ SYK model with $\qq=3$, for $N_c=3,4,\dots,17$. The symmetry classes are
\raisebox{2pt}{\colorbox{cGOE}{\quad\tiny$\strut$\!\!\!}}\,:~GOE, 
\raisebox{2pt}{\colorbox{cGSE}{\quad\tiny$\strut$\!\!\!}}\,:~GSE, and uncolored numbers GUE.
All of these results were checked numerically.%
\footnote{Our tables are correct ``almost surely'', i.e., there can be deviations 
from the numbers in the tables if the random couplings $\{X_{ijk}\}$ in 
\eqref{eq:qqbardef} are fine-tuned (e.g., to all zeros). 
Such exceptional cases are of measure zero and physically unimportant.} 
\setlength{\tabcolsep}{3pt}
\\
\begin{minipage}{.2\textwidth}
	\noindent 
	\paragraph{$\blacksquare$~$N_c=3$}~
	\\*[7pt]
	\mbox{\quad} 
	\begin{tabular}{c|cccc}
		$f$ & 0 &1&2&3
		\\\hline 
		$N_f^+$ &1&0&0&0
		\\
		$N_f^-$ &0&0&0&1
		\\
		$N_f^z$ &0&3&3&0
	\end{tabular}
\end{minipage}
\begin{minipage}{.23\textwidth}
	\noindent 
	\paragraph{$\blacksquare$~$N_c=4$}~
	\\*[7pt]
	\mbox{\quad} 
	\begin{tabular}{c|ccccc}
		$f$ & 0 &1&2&3&4
		\\\hline 
		$N_f^+$ &1&1&0&0&0
		\\
		$N_f^-$ &0&0&0&1&1
		\\
		$N_f^z$ &0&3&6&3&0
	\end{tabular}
\end{minipage}
\begin{minipage}{.26\textwidth}
	\noindent 
	\paragraph{$\blacksquare$~$N_c=5$}~
	\\*[7pt] 
	\mbox{\quad}
	\begin{tabular}{c|cccccc}
		$f$ &0&1&2&3&4&5 
		\\\hline 
		$N_f^+$ &1&\cellcolor{cGSE}4&1&0&0&0
		\\
		$N_f^-$ &0&0&0&1&\cellcolor{cGSE}4&1
		\\
		$N_f^z$ &0&1&9&9&1&0
	\end{tabular}
\end{minipage}
\begin{minipage}{.29\textwidth}
	\noindent 
	\paragraph{$\blacksquare$~$N_c=6$}~
	\\*[7pt]
	\mbox{\quad}
	\begin{tabular}{c|ccccccc}
		$f$ &0&1&2&3&4&5&6
		\\\hline 
		$N_f^+$ &1&6&6&1&0&0&0
		\\
		$N_f^-$ &0&0&0&1&6&6&1
		\\
		$N_f^z$ &0&0&9&18&9&0&0
	\end{tabular}
\end{minipage}
\\*[7pt]
\begin{minipage}{.43\textwidth}
	\noindent 
	\paragraph{$\blacksquare$~$N_c=7$}~
	\\*[7pt]
	\mbox{\quad}
	\begin{tabular}{c|cccccccc}
		$f$ &0&1&2&3&4&5&6&7
		\\\hline 
		$N_f^+$ &1&7&\cellcolor{cGOE}21&7&1&0&0&0
		\\
		$N_f^-$ &0&0&0&1&7&\cellcolor{cGOE}21&7&1
		\\
		$N_f^z$ &0&0&0&27&27&0&0&0
	\end{tabular}
\end{minipage}
\begin{minipage}{.45\textwidth}
	\noindent 
	\paragraph{$\blacksquare$~$N_c=8$}~
	\\*[7pt]
	\mbox{\quad}
	\begin{tabular}{c|ccccccccc}
		$f$ &0&1&2&3&4&5&6&7&8
		\\\hline 
		$N_f^+$ &1&8&28&28&8&1&0&0&0
		\\
		$N_f^-$ &0&0&0&1&8&28&28&8&1
		\\
		$N_f^z$ &0&0&0&27&54&27&0&0&0
	\end{tabular}
\end{minipage}
\\*[7pt]
\begin{minipage}{.427\textwidth}
	\noindent 
	\paragraph{$\blacksquare$~$N_c=9$}~
	\\*[7pt]
	\mbox{\quad}
	\begin{tabular}{c|cccccccccc}
		$f$ &0&1&2&3&4&5&6&7&8&9
		\\\hline 
		$N_f^+$ &1&9&36&\cellcolor{cGSE}80&36&9&1&0&0&0
		\\
		$N_f^-$ &0&0&0&1&9&36&\cellcolor{cGSE}80&36&9&1
		\\
		$N_f^z$ &0&0&0&3&81&81&3&0&0&0
	\end{tabular}
\end{minipage}
\begin{minipage}{.567\textwidth}
	\noindent 
	\paragraph{$\blacksquare$~$N_c=10$}~
	\\*[7pt]
	\mbox{\quad}
	\begin{tabular}{c|ccccccccccc}
		$f$ &0&1&2&3&4&5&6&7&8&9&10
		\\\hline 
		$N_f^+$ &1&10&45&119&119&45&10&1&0&0&0
		\\
		$N_f^-$ &0&0&0&1&10&45&119&119&45&10&1
		\\
		$N_f^z$ &0&0&0&0&81&162&81&0&0&0&0
	\end{tabular}
\end{minipage}
\paragraph{$\blacksquare$~$N_c=11$}~
\\*[7pt]
\mbox{\quad}
\begin{tabular}{c|cccccccccccc}
	$f$ &0&1&2&3&4&5&6&7&8&9&10&11
	\\\hline 
	$N_f^+$ &1&11&55&164&\cellcolor{cGOE}319&164&55&11&1&0&0&0
	\\
	$N_f^-$ &0&0&0&1&11&55&164&\cellcolor{cGOE}319&164&55&11&1
	\\
	$N_f^z$ &0&0&0&0&0&243&243&0&0&0&0&0
\end{tabular}
\paragraph{$\blacksquare$~$N_c=12$}~
\\*[7pt]
\mbox{\quad}
\begin{tabular}{c|ccccccccccccc}
	$f$ &0&1&2&3&4&5&6&7&8&9&10&11&12
	\\\hline 
	$N_f^+$ &1&12&66&219&483&483&219&66&12&1&0&0&0
	\\
	$N_f^-$ &0&0&0&1&12&66&219&483&483&219&66&12&1
	\\
	$N_f^z$ &0&0&0&0&0&243&486&243&0&0&0&0&0
\end{tabular}
\paragraph{$\blacksquare$~$N_c=13$}~
\\*[7pt]
\mbox{\quad}
\begin{tabular}{c|cccccccccccccc}
	$f$ &0&1&2&3&4&5&6&7&8&9&10&11&12&13
	\\\hline 
	$N_f^+$ &1&13&78&285&702&\cellcolor{cGSE}1208&702&285&78&13&1&0&0&0
	\\
	$N_f^-$ &0&0&0&1&13&78&285&702&\cellcolor{cGSE}1208&702&285&78&13&1
	\\
	$N_f^z$ &0&0&0&0&0&1&729&729&1&0&0&0&0&0
\end{tabular}
\paragraph{$\blacksquare$~$N_c=14$}~
\\*[7pt]
\mbox{\quad}
\begin{tabular}{c|ccccccccccccccc}
	$f$ &0&1&2&3&4&5&6&7&8&9&10&11&12&13&14
	\\\hline 
	$N_f^+$ &1&14&91&363&987&1911&1911&987&363&91&14&1&0&0&0
	\\
	$N_f^-$ &0&0&0&1&14&91&363&987&1911&1911&987&363&91&14&1
	\\
	$N_f^z$ &0&0&0&0&0&0&729&1458&729&0&0&0&0&0&0
\end{tabular}
\paragraph{$\blacksquare$~$N_c=15$}~
\\*[7pt]
\mbox{\quad}
\begin{tabular}{c|cccccccccccccccc}
	$f$ &0&1&2&3&4&5&6&7&8&9&10&11&12&13&14&15
	\\\hline 
	$N_f^+$ &1&15&105&454&1350&2898&\cellcolor{cGOE}4551&2898&1350&454&105&15&1&0&0&0 
	\\
	$N_f^-$ &0&0&0&1&15&105&454&1350&2898&\cellcolor{cGOE}4551&2898&1350&454&105&15&1
	\\
	$N_f^z$ &0&0&0&0&0&0&0&2187&2187&0&0&0&0&0&0&0
\end{tabular}
\paragraph{$\blacksquare$~$N_c=16$}~
\\*[7pt]
\mbox{\quad}\scalebox{.987}
{\begin{tabular}{c|ccccccccccccccccc}
	$f$ &0&1&2&3&4&5&6&7&8&9&10&11&12&13&14&15&16
	\\\hline 
	$N_f^+$ &1&16&120&559&1804&4248&7449&7449&4248&1804&559&120&16&1&0&0&0
	\\
	$N_f^-$ &0&0&0&1&16&120&559&1804&4248&7449&7449&4248&1804&559&120&16&1
	\\
	$N_f^z$ &0&0&0&0&0&0&0&2187&4374&2187&0&0&0&0&0&0&0
\end{tabular}}
\paragraph{$\blacksquare$~$N_c=17$}~
\\*[7pt]
\mbox{\quad}\scalebox{0.864}
{\begin{tabular}{c|cccccccccccccccccc}
	$f$ &0&1&2&3&4&5&6&7&8&9&10&11&12&13&14&15&16&17
	\\\hline 
	$N_f^+$ &1&17&136&679&2363&6052&11697&\cellcolor{cGSE}17084&11697&6052&2363&679&136&17&1&0&0&0
	\\
	$N_f^-$ &0&0&0&1&17&136&679&2363&6052&11697&\cellcolor{cGSE}17084&11697&6052&2363&679&136&17&1
	\\
	$N_f^z$ &0&0&0&0&0&0&0&1&6561&6561&1&0&0&0&0&0&0&0
\end{tabular}}

\bibliography{SYK_manuscript_for_JHEP_v3.bbl}
\end{document}